\documentclass[conference]{IEEEtran}

\newcommand{\sys}{NFSlicer\xspace}

\newcommand{\ie}{{i.e.,}\xspace}

\newcommand{\etal}{{et al.}\xspace}

\newcommand{\microsecond}{\ensuremath{\upmu{}}s\xspace}

\newcommand{\iscanew}[1]{{\color{blue} #1}}

\newcommand{\squishitemize}{
 \begin{list}{$\bullet$}
  { \setlength{\itemsep}{0pt}
     \setlength{\parsep}{3pt}
     \setlength{\topsep}{0pt}
     \setlength{\partopsep}{0pt}
     \setlength{\leftmargin}{1.95em}
     \setlength{\labelwidth}{1.5em}
     \setlength{\labelsep}{0.5em} } }

\newcounter{Lcount}
\newcommand{\squishlist}{
    \begin{list}{\arabic{Lcount}. }
   { \usecounter{Lcount}
        \setlength{\itemsep}{0pt}
        \setlength{\parsep}{3pt}
        \setlength{\topsep}{0pt}
        \setlength{\partopsep}{0pt}
        \setlength{\leftmargin}{2em}
        \setlength{\labelwidth}{1.5em}
        \setlength{\labelsep}{0.5em} } }

\newcommand{\squishend}{\end{list}}

\usepackage[normalem]{ulem}
\usepackage{xspace}
\usepackage{listings}
\usepackage{comment}
\usepackage{xcolor}
\usepackage{color}
\usepackage{graphicx}
\usepackage{subfig}
\usepackage{upgreek}
\usepackage{multirow}

\usepackage[bookmarks=true,breaklinks=true,letterpaper=true,colorlinks,linkcolor=black,citecolor=blue,urlcolor=black]{hyperref}

\definecolor{mGreen}{rgb}{0,0.6,0}
\definecolor{mGray}{rgb}{0.5,0.5,0.5}
\definecolor{mPurple}{rgb}{0.58,0,0.82}
\definecolor{backgroundColour}{rgb}{0.95,0.95,0.92}

\lstdefinestyle{CStyle}{
    backgroundcolor=\color{backgroundColour},   
    commentstyle=\color{mGreen},
    keywordstyle=\color{magenta},
    numberstyle=\tiny\color{mGray},
    stringstyle=\color{mPurple},
    basicstyle=\footnotesize,
    breakatwhitespace=false,         
    breaklines=true,                 
    captionpos=b,                    
    keepspaces=true,                 
    numbers=left,                    
    numbersep=5pt,                  
    showspaces=false,                
    showstringspaces=false,
    showtabs=false,                  
    tabsize=2,
    language=C
}

\usepackage{balance}
\usepackage[capitalize,noabbrev,nameinlink]{cleveref}
\crefformat{section}{\S#2#1#3} 
\crefformat{subsection}{\S#2#1#3}
\crefformat{subsubsection}{\S#2#1#3}

\Crefname{figure}{Fig.}{Figs.}
\Crefname{equation}{Eqn.}{Eqns.}
\Crefname{section}{\S}{\S}

\begin{document}

\title{
\sys{}: 
Data Movement Optimization\\ for Shallow Network Functions }

\date{}

\author{Anirudh Sarma\quad Hamed Seyedroudbari \quad Harshit Gupta \quad Umakishore Ramachandran \quad Alexandros Daglis\\\textit{Georgia Institute of Technology}}

\maketitle
\thispagestyle{plain}
\pagestyle{plain}

\begin{abstract}

Network Function (NF) deployments on commodity servers have become ubiquitous in datacenters and enterprise settings. Many commonly used NFs such as firewalls, load balancers and NATs are shallow---i.e., they only examine the packet's header, despite the entire packet being transferred on and off the server. As a result, the gap between moved and inspected data when handling large packets exceeds $20\times$. At modern network rates, such excess data movement is detrimental to performance, hurting both the average and 90\% tail latency of large packets by up to $1.7\times$. Our thorough performance analysis identifies high contention on the NIC-server PCIe interface and in the server's memory hierarchy as the main bottlenecks.

We introduce \sys, a data movement optimization implemented as a NIC extension to mitigate the bottlenecks stemming from data movement deluge in deployments of {shallow} NFs {on commodity servers}.
\sys only transfers the small portion of each packet that the deployed NFs actually inspect, by \textit{slicing} the packet's payload and temporarily storing it in on-NIC memory. When the server later transmits the processed packet, \sys \textit{splices} it to its previously sliced payload. We develop a software-based emulation platform and demonstrate that \sys effectively minimizes data movement between the NIC and the server, bridging the latency gap between small and large packet NF processing. On a range of shallow NFs handling 1518B packets, \sys reduces average and 90\% tail latency by up to 17\% / 29\%, respectively.

\end{abstract}
\section{Introduction}
\label{sec:intro}

Network Functions (NFs) have been transitioning from specialized middleboxes to virtualized counterparts on commodity servers \cite{katsikas:metron, koponen:virtualization,  martins:clickos, sun:nfp}, which increases their flexibility and facilitates their deployment, boosting their ubiquity in enterprise and datacenter settings \cite{sherry:making}.
The vast majority of internet traffic that targets an online service is first filtered through several NFs deployed on general-purpose servers, offering functionality such as load balancing within the datacenter, and applying forwarding and firewall rules. 
Due to the latency-sensitive nature of modern online services, the latency incurred on each packet's NF processing is an important figure of merit.

NFs as a workload are extremely network-bandwidth intensive, as they typically involve minimal processing per received packet. Modern high-bandwidth NICs are therefore a welcome addition to NF-handling servers, as they enable better resource consolidation in the datacenter.
However, a side-effect of increasing line rates is unprecedented network traffic moving on and off the server, promoting data movement to a first-order performance determinant for latency-sensitive NFs. 

\begin{figure}
    \centering
    \includegraphics[width=0.6\linewidth]{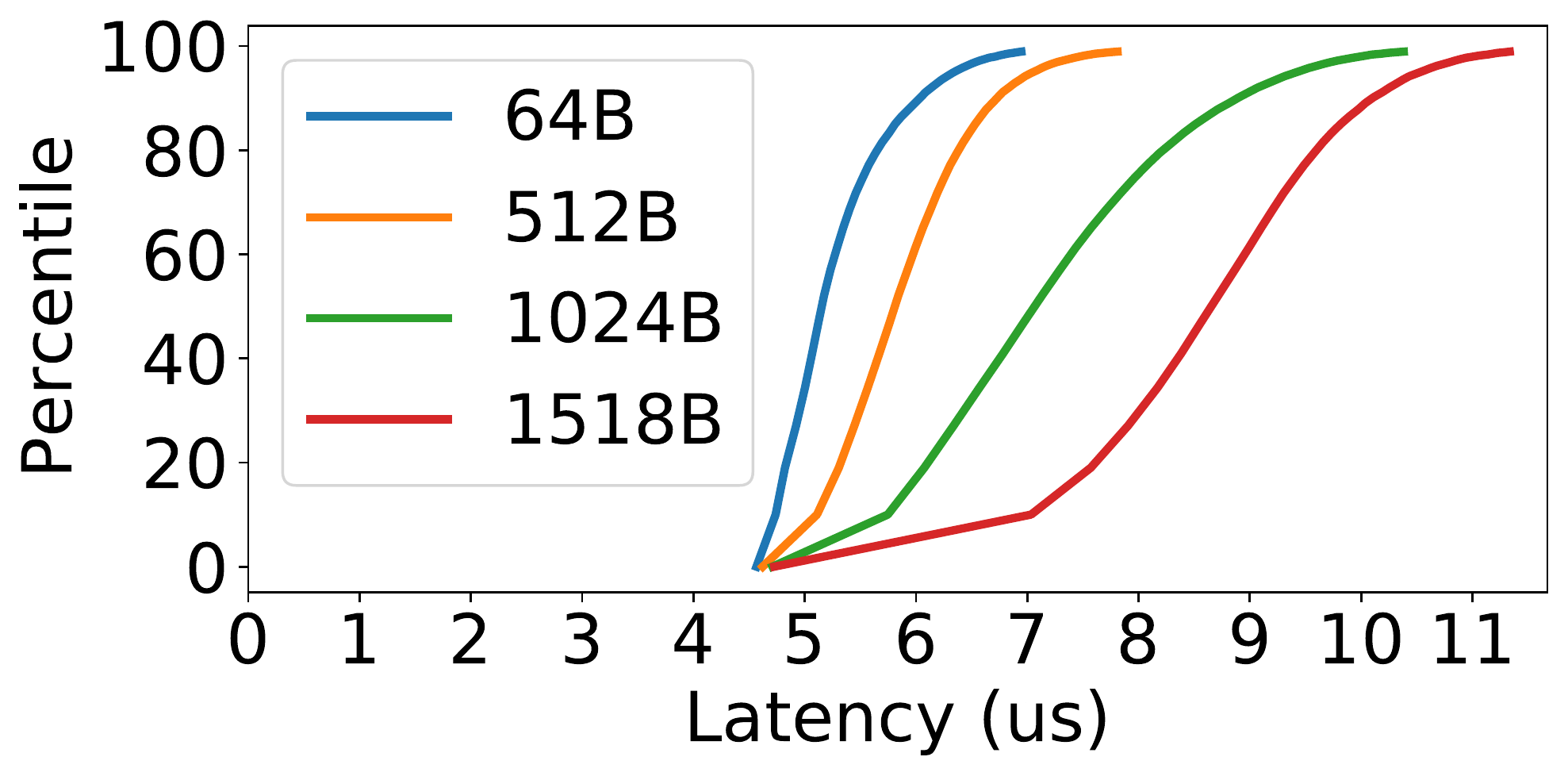}
    \caption{L2 forwarder NF latency CDF for variable size packets at 7Mpps arrival rate, measured on directly connected client-server machines equipped with 100G NIC (details in \cref{sec:method}).} 
    \label{fig:motivation}
    \vspace{-4mm}
\end{figure}

\cref{fig:motivation} exemplifies the impact of network data movement, by showing the end-to-end response latency distribution for an L2 forwarder NF handling network traffic of four different packet sizes.
While the base latency for all packet sizes is the same ($\sim$4.5\microsecond), the latency gap quickly grows as a function of packet size to $1.7\times$ between 1518B and 64B packets.
Although the latency difference may initially seem intuitive, L2 forwarders, like \textit{many NFs used in production, are shallow}---i.e., they only operate on packet headers, hence processing cost is insensitive to packet size.
We therefore posit that the observed latency gap is primarily attributed to redundant data movement.
Given the deluge of large-packet network traffic---video will account for 82\% of internet traffic by 2022 \cite{cisco:video}---mitigating inefficiencies in its handling is critical.

In this work, we propose \sys, {a data movement optimization implemented as} a NIC extension to mitigate the latency overheads encountered by shallow NF when handling large packets.
\sys is a software-hardware co-design that minimizes data movement between a server executing shallow NFs and its NIC, by only transferring the small subset of the packet the NFs need.
{The core of \sys is a new basic on-NIC operation, called packet  \textit{Slice \& Splice}:}
as each network packet bound to be processed by a shallow NF arrives from the network, NFSlicer \textit{slices} the packet's payload,  temporarily stores it in on-NIC memory resources, and extends the packet's header with special metadata.
After a CPU on the server executes the NF of the received header and transmits it back to the network, NFSlicer uses these metadata to locate the outgoing packet's corresponding payload and \textit{splices} it back to the header to reconstruct the full packet before placing it on the wire.
{Slice \& Splice addresses large-packet handling inefficiency at its source}, shrinking server-NIC data transfers by more than $20\times$ for large packets.  

The basic idea of payload slicing for shallow NFs was recently introduced by Goswami \etal{}'s PayloadPark~\cite{goswami:parking}.
Although \sys's design bears strong similarity with PayloadPark, our contributions are distinct.
PayloadPark employs a \textit{partial} Slice \& Splice approach on a programmable switch to improve link goodput.
In contrast, \sys demonstrates, for the first time, (i) the system-level effects of redundant on-server network data movement; and (ii) the performance gain potential that is \textit{only} attainable by eliminating it with \textit{full} payload slicing, which is beyond the capabilities of switch-based solutions.
We further show that \sys's NIC-based mechanism is fundamentally more scalable than switch-based approaches.
Overall, while PayloadPark and \sys share significant conceptual similarities, they are complementary mechanisms with distinct strengths. 

In summary, we make the following contributions:
\begin{itemize}
    \item We conduct a thorough microarchitectural study and identify that the noticeable latency gap between large- and small-packet NF processing is attributed to redundant data movement. Our analysis identifies the primary bottleneck on the PCIe interface, and on memory bandwidth to a lesser extent. Thus, the only solution to bridge that latency gap is to reduce the amount of data moved.
    \item We design the protocol and hardware extensions necessary to realize NFSlicer as a NIC extension. 
    \item We develop a software emulation platform to evaluate NFSlicer's performance improvement potential for a range of shallow NFs. Under the throughput limitations of our emulation platform, we show that for MTU-size (\ie 1500B) packets, NFSlicer improves the median and 90\% tail latency of shallow NFs by 17--20\% and 9--29\%, respectively. We further show that for higher packet rates, the tail latency improvement potential grows to 55\%.
    \item We synthesize a hardware implementation of NFSlicer to quantify the resource needs and added latency of the Slice \& Splice operation,  {demonstrating \sys's feasibility.} 
\end{itemize}

The rest of the paper is organized as follows. \cref{sec:background} provides brief background on NFs, typical network packet sizes, and modern NIC capabilities. \cref{sec:design} and \cref{sec:impl} describe \sys's design and our corresponding emulation platform implementation, respectively. We detail our methodology in \cref{sec:method} and evaluate \sys in \cref{sec:eval}.
We present a hardware implementation and our synthesis results in \cref{sec:hardware-impl}.
Finally, \cref{sec:discussion} discusses limitations and potential extensions, 
\cref{sec:relwork} covers related work, 
and \cref{sec:conclusion} concludes.

\section{Background}
\label{sec:background}

\subsection{Network Functions and Modern NF Deployment}
NFs broadly range from  network architecture controllers implementing SDN control functionality to simple networking utilities such as software  switches, routers, Network  Address  Translators (NAT), firewalls, intrusion  detection  systems, load balancers, WAN optimizers and flow monitors \cite{li:nfvsurvey}.
Several of these NFs such as firewalls, load balancers, NAT, and switches are ``shallow''---i.e., they do not require the entire packet for processing, as they only inspect and modify the packet's L2--L4 headers. 
\sys's data movement optimization is directly applicable to all such shallow NFs.

{Historically, NFs were usually deployed on specialized middleboxes. 
However, the recent trend of NF virtualization has triggered a major shift from such middleboxes to commodity off-the-shelf servers.
The reason for this transition is not performance, but rather ease of deployment and improved resource consolidation in multi-tenant cluster environments.}

\subsection{Large-Packet Dominance of Internet Traffic}

\sys is an optimization specifically targeting large network packets, which dominate Internet traffic. 
A large contributor to this trend is the growing demand for video, in the form of IPTV, media streaming, surveillance, etc. To illustrate, video accounted for 58\% of total internet traffic in 2018 \cite{statista:video} and is expected to grow to 82\% by 2022 \cite{cisco:video}. 
We empirically confirm the prevalence of large-packet network traffic by crunching a packet capture from an Internet backbone link provided by the Center for Applied Internet Data Analysis \cite{caida:internet-traffic}.
We find that 49\% of all packets are at least 1400B and 57\% are larger than 500B.
More importantly, large packets completely dominate the data volume moved on the Internet: 84\% / 93\% of total data volume is attributed to packets $\geq$1400B / $\geq$500B, respectively.
As most of this traffic is routed through several NFs deployed on commodity servers, \sys's optimization for large-packet NF processing has broad applicability.

\subsection{Advanced NICs}

In the last few years, we have witnessed an explosion in NIC capabilities, both in terms of added functionality and/or programmability. It is now common for modern NICs to offer specialized hardware for networking functionality offload, such as checksum computations and encryption.
As \textit{what} we should be accelerating on NICs is still unclear, a second growing trend 
is the provisioning of programmable hardware resources on the NIC \cite{kaufmann:flexnic, mpr:bluefield, innova,  putnam:reconfigurable}.

Two technological trends are contributing to the evolution of NICs in these directions. 
First, the end of Moore's Law is pushing architects toward hardware specialization, which is fueling an appetite for in-network computing \cite{sapio:dumb-idea}. 
Architects have found renewed interest in the quest for networking and application functionality that can be moved from the CPU to specialized hardware residing in network gear, such as switches and NICs.
Second, increasing line rates (commercial NICs can already drive 2x200G of network traffic \cite{mellanox:connectx6}) require more pins, which in turn require larger dies and more available on-NIC silicon that can be leveraged to offer additional functionality.  
Given the momentum for NIC evolution and specialization, \sys is an appealing NIC extension promising significant latency gains for an important workload class that can be realistically deployed in production environments in the near future.
{We consider a hardened IP block as the most fitting implementation of \sys's Splice \& Splice operation, as it is a basic primitive that can be leveraged across shallow NFs. However, in the near term, a soft form of the operation can also be implemented on a NIC with programmable hardware resources.}
\section{\sys Design}
\label{sec:design}

We begin this section with \sys's design overview, elaborate on the software-hardware interface to enable packet slicing, and then describe the Slice \& Splice pipeline.

\subsection{Overview}

\begin{figure}[t]
    \centering
    \includegraphics[width=.9\linewidth]{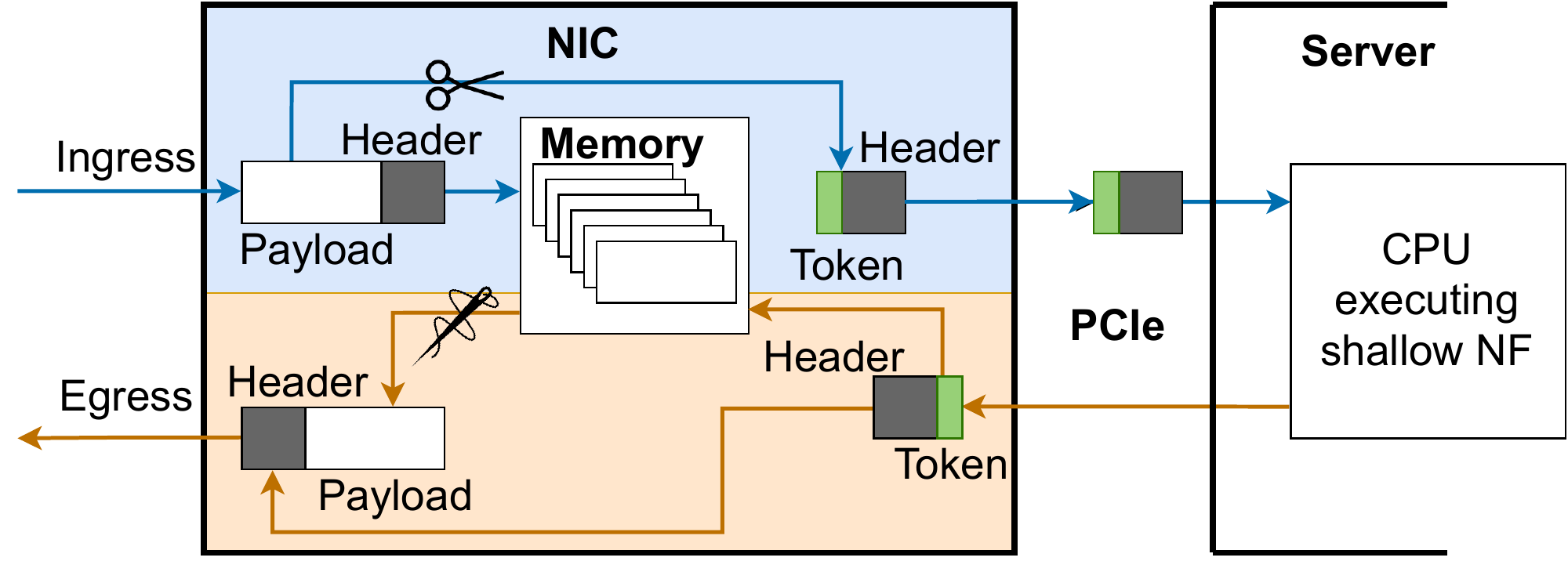}
    \caption{\sys overview.}
	\label{fig:overview}
\end{figure}

\sys is a NIC extension designed to avoid moving a packet's payload between a server and its NIC, by temporarily storing it in on-NIC memory resources.
\cref{fig:overview} displays a high-level view of \sys's operation.
As a full packet arrives from the network, the NIC \textit{slices} its \textit{entire} payload and stores it in local memory.
The payload is replaced by a small \textit{token}, which serves as a unique identifier to retrieve the sliced payload later, on the packet's egress path. The transformed packet is then transferred to the server, where the server applies its shallow NF processing and transmits it back to the NIC.
On the packet's egress, \sys uses the embedded token to retrieve the packet's corresponding payload and \textit{splices} it to the processed header, before transmitting the reconstructed packet on the wire.
As pointed out in \cref{sec:intro}, the design of \sys's mechanism is similar to Goswami \etal{}'s PayloadPark~\cite{goswami:parking}, with {the following key differences: (i) \sys targets a NIC-based implementation rather than a switch-based implementation that has to conform to fundamental limitations associated with RMT (Reconfigurable Match Table) pipelines; and (ii) instead of \textit{partial} payload slicing to improve network link goodput, \sys enables \textit{entire} payload slicing to eliminate data movement bottlenecks between the NIC and the server.
}

\subsection{Software-Hardware Interface}
\label{sec:design:interface}

\begin{figure}
    \centering
    \includegraphics[width=\linewidth]{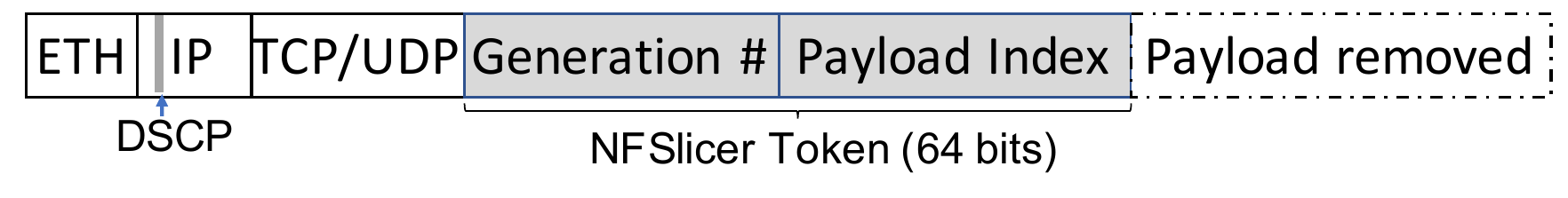}
    \vspace{-6mm}
    \caption{Packet layout after being processed by \sys.}
    \label{fig:packet-layout}
    \vspace{-3mm}
\end{figure}

\cref{fig:packet-layout} shows \sys's software-hardware interface.
We use the IP header's DSCP field, by setting it to a special value to mark packets that have been sliced.
We use the value $0b111111$, which is reserved for experimental/local use \cite{dscp}.
When \sys has available resources to slice and buffer an incoming packet's payload, it sets the DSCP special value, removes the packet's payload, and extends its header with the \textit{\sys token,} a 64-bit value that uniquely identifies a sliced packet's corresponding payload.
The token is later used on the packet's egress path---after it has been processed by the NF on the server---to retrieve the packet's corresponding payload for reconstruction.
\cref{sec:design:ops} details the purpose of the token's two fields---\textit{payload index} and \textit{generation number}.

\subsection{Slice \& Splice Operation}
\label{sec:design:ops}

\begin{figure}[t]
    \centering
    \includegraphics[width=\linewidth]{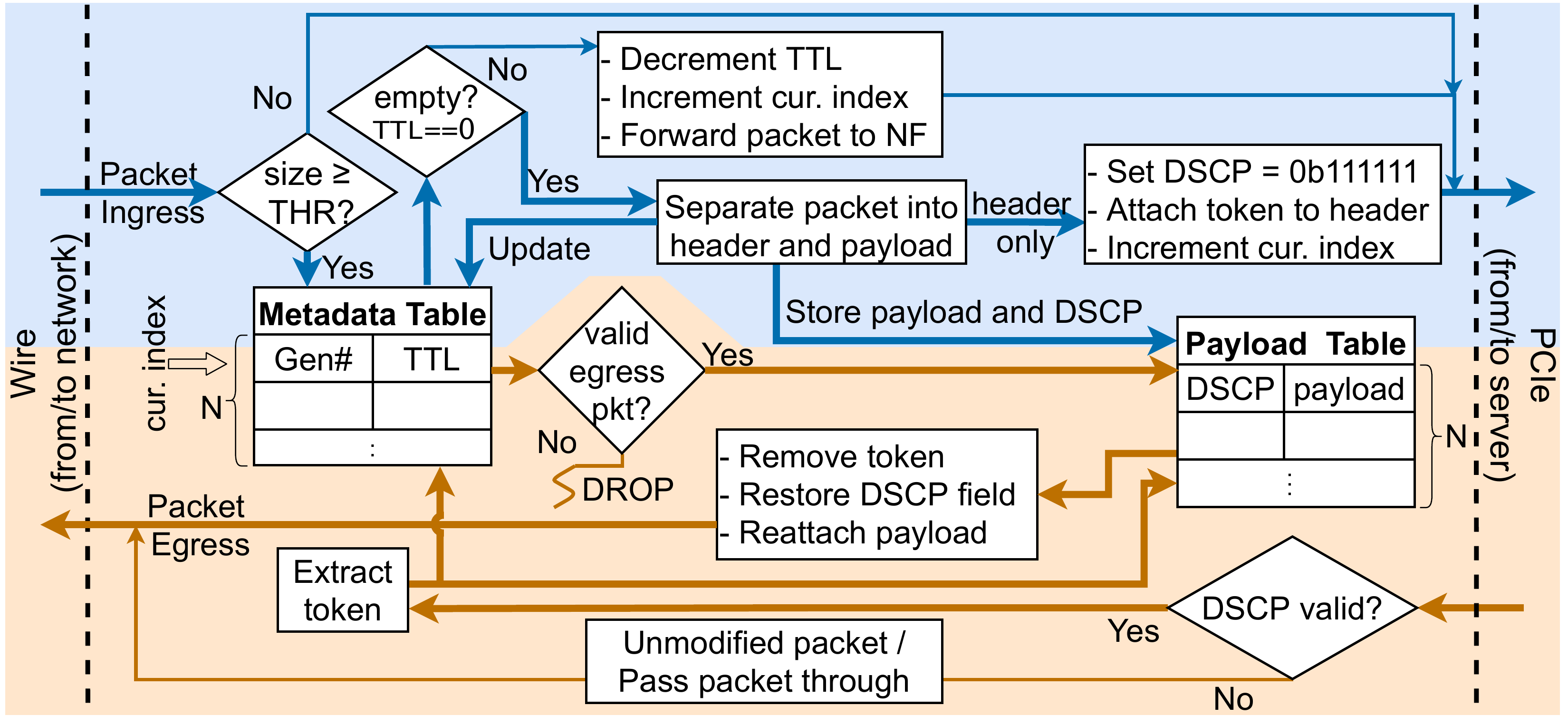}
    \caption{Slice \& Splice pipeline. Top half: slice op; bottom half: splice op.  Thick lines highlight the common path.}
		\label{fig:slice-splice-pipeline}
\end{figure}

\cref{fig:slice-splice-pipeline} shows the steps involved in \sys's Slice \& Splice operations. 
\sys uses the \textit{payload} table to store the sliced payloads, and a corresponding \textit{metadata} table to keep track of occupancy and ensure correctness while splicing. 
Each table has $N$ entries and an index points to the next available entry. 
$N$ is provisioned to comfortably accommodate the bandwidth-delay product of maximum arrival rate of 
\textit{slice-worthy} packets  and the maximum expected average service time of the target deployment's shallow NFs.
\textit{Slice-worthy} denotes the size below which slicing yields diminishing performance gains.
For a target 100G system, we experimentally found this value to be 500B, which we denote as \textit{threshold} (THR).

Table entries are allocated in FIFO order---i.e., the index is incremented after each arrival of a packet of size $\geq$ THR. Each entry in the metadata table comprises a generation number and a Time-To-Live (TTL) field.
TTL enables a low-cost stale-entry garbage collection mechanism; {as soon as a stored payload's TTL hits zero, the entry is discarded.}
The generation number is used to verify---on a packet's egress---that the payload retrieved from the payload table is a correct match. 

\paragraph{Slice operation}

The thick lines in \cref{fig:slice-splice-pipeline}'s top half show the workflow in \sys's slice operation for the common case.
When a packet arrives on the ingress path, \sys first checks if the packet is large enough to justify slicing (size $\geq$ THR). 
If so, \sys creates a token containing the current index of the entry where the payload will be placed, and increments this index to point to the next empty entry. 
It then splits the packet into its corresponding header and payload, and stores the payload and DSCP field in the \textit{payload} table. 
Finally, \sys marks that the packet has been sliced by setting the packet header's DSCP field to $0b111111$ and extends the header with the token before sending it to the server for NF Processing. 
{A packet may be forwarded to the server without getting sliced for two reasons: 
(i) it is not slice-worthy (i.e., size $<$ THR); or
(ii) the current index does not point to an available entry---i.e., not finding readily available space in \sys's structures does \textit{not} cause a packet drop.
}

\paragraph{Splice operation}

The thick lines in \cref{fig:slice-splice-pipeline}'s bottom half  show the workflow in \sys's splice operation for the common case.
Once the packet arrives from the server for transmission, \sys checks the DSCP field to confirm if the packet was previously sliced. A non-sliced packet is transmitted out on the wire without further steps. 
For sliced packets, \sys removes and parses the token to obtain the index to the stored payload, restores the packet's original DSCP field and payload, transmits the reconstructed packet to the network, and resets the metadata table's  entry to zero.

\paragraph{Payload timeouts}

NFs may explicitly drop packets as part of their operation (e.g., block rule of a firewall). \sys thus requires a garbage collection mechanism to reclaim entries of dropped packets. 
We use the TTL field to set a validity duration for each metadata/payload table entry. 
A new packet on the ingress path that finds the current index pointing to an entry with a non-zero TTL is forwarded to the server for NF processing without getting sliced.
\sys decrements the pointed entry's TTL field and advances the current index.
As the table index operates in FIFO order and each entry's TTL is decremented on each access, the TTL represents the number of allowable table wrap-arounds before an entry is evicted.
{In rare events of extreme processing delay spikes on the server, an outgoing sliced packet may not find its corresponding stored in the payload table, in which case the packet is dropped. 
Such occurrences indicate transient system overload conditions, and dropping packets has been recently used as a mechanism for improved performance predictability and overload control for microsecond-scale latency-sensitive services \cite{cho:overload,sutherland:nebula,tootoonchian:resq}.
Therefore, \sys's rare TTL-induced packet drops will be masked by throttling already happening at higher levels of the network stack.
Besides, latency-sensitive applications, like NFs, typically have a maximum acceptable tail latency to preserve QoS (e.g., $10\times$ of average service time). The TTL can be configured to accommodate the application’s acceptable tail latency, so that any forced packet drops only occur for packets that are way overdue their acceptable response time. For example, given that the base buffering capacity is provisioned to accommodate for the expected average service time, setting TTL to 10 would allow 10 buffer wraparounds, hence allowing \textit{at least} $10\times$ average service time residency. 
}

\paragraph{Payload retrieval correctness} 
{As shown in \cref{fig:packet-layout}, the \sys token consists of two fields:}
the 64-bit token's lowest-order $\log_2{N}$ bits encode the payload index, the remaining bits encode a generation number, 
{which is required to guarantee correct payload splicing on a packet's egress.}
The generation number is unique per entry and is incremented whenever a new entry is inserted into the metadata/payload table. On the egress path, the generation number encoded in the outgoing packet's token is matched against the generation number stored in the metadata tables to prevent erroneous splicing of a newer packet's payload to an older outgoing packet's header, {a rare situation that arises when an entry's TTL has expired and a delayed response packet corresponding to that evicted entry is transmitted from the server}.

\section{Implementation}
\label{sec:impl}

\begin{figure}[t]
    \centering
    \includegraphics[width=\linewidth]{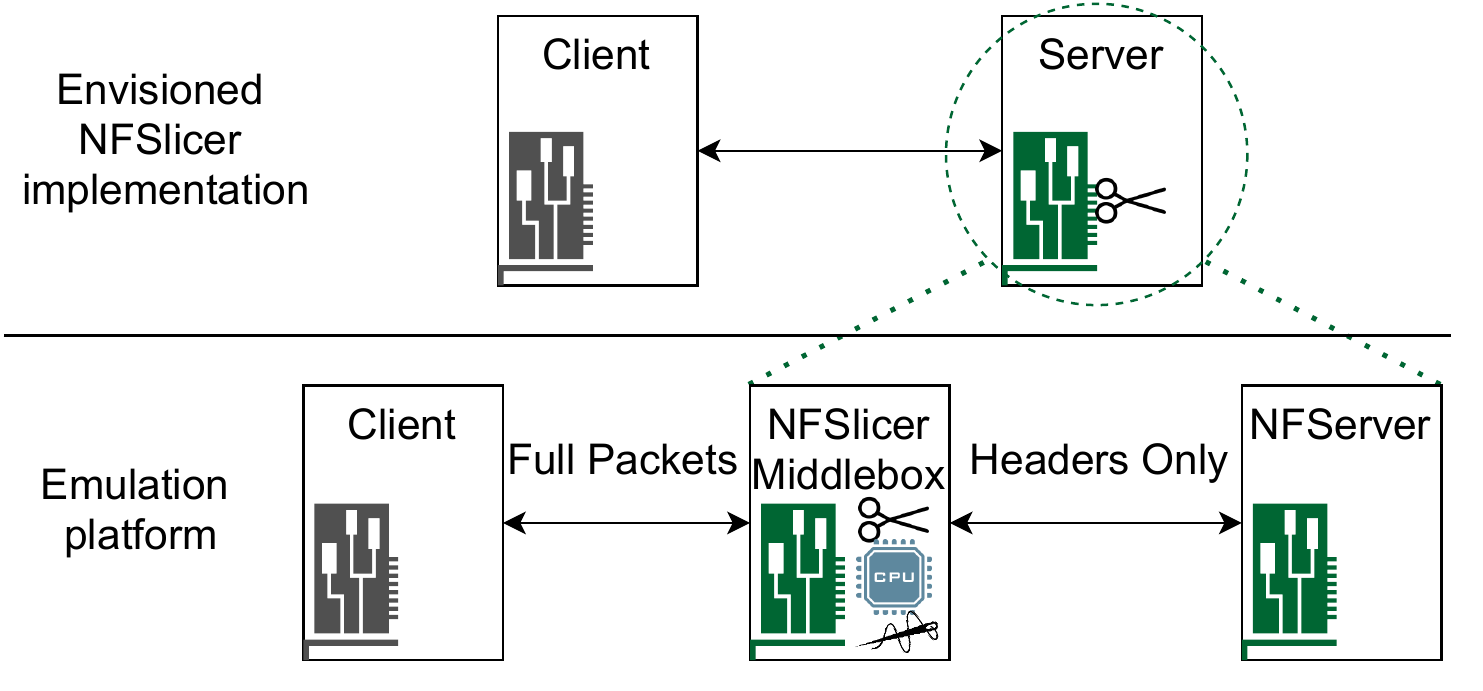}
        
    \caption{\sys software emulation platform developed as a DPDK application. 
    }

    \label{fig:emulation}
\end{figure}

We envision \sys as a hardware mechanism implemented on a NIC. 
In this work, we  evaluate the Slice \& Splice technique's performance effect by developing a software emulation platform, which allows us to accomplish three significant goals without undergoing the considerable engineering effort of developing a fully functional hardware prototype:
(i) verify the protocol's functional correctness, 
(ii) evaluate the technique's performance improvement \textit{potential}, and 
(iii) study the microarchitectural bottlenecks on the NF server when handling packets of different sizes. 
We emulate the envisioned NFSlicer functionality of the NIC in software, on a separate server.
\cref{fig:emulation} shows how the emulation platform decomposes the envisioned NFSlicer-enabled server into two servers, directly connected with a 100G link: a \textit{``middlebox''} and an \textit{``NFServer''}. 
All traffic between clients and the NFServer flows through the middlebox, which performs the Slice \& Splice functionality. 
The NFServer receives sliced instead of full-size packets of the  client-originating packet flows, but executes unmodified shallow NFs.

The middlebox's Slice \& Splice operations are implemented as a DPDK-based NF, which is performed in a run-to-completion fashion: a packet is received, processed and transmitted by a core before retrieving the next packet. A CPU core polls for packet arrivals on a specific port's \textit{RX} descriptor ring, performs the Slice or Splice operation, and enqueues the transformed packet in the transmit queue.
We maintain per-core arrays of \texttt{payload\_t} and \texttt{metadata\_t} structures (\cref{lst:emustructs}) to store payloads and their corresponding metadata, respectively. Array sizes are specified at initialization and are provisioned to sustain the emulation platform's peak NF processing bandwidth-delay product.
We further discuss array size provisioning for a hardware-based \sys implementation, where buffering capacity is constrained by hardware limitations, in \cref{sec:hardware-impl}. 
The middlebox performs a Slice or a Splice operation on each incoming packet, depending on its direction: client-to-server \textit{(ingress)}, or server-to-client \textit{(egress).}

\paragraph{Slice operation}
For each packet received on the ingress path, the middlebox determines the packet's payload size that must be sliced and stored in memory in order to forward the smallest possible packet to the NFServer. 
The middlebox maintains an index into the \texttt{payloads} array to store the payload after slicing. If the \texttt{metadata\_table} is found full, the packet is forwarded to the NFServer unmodified.  
Otherwise, the middlebox sets the \texttt{ttl} to a predefined threshold 
and increments the generation number in the \texttt{metadata\_table}. 
After storing the payload and the DSCP field in the \texttt{payloads} array, the middlebox appends a \texttt{token} to the packet containing the payload's index in the array and the generation number.

\paragraph{Splice operation}
For each packet received on the egress path, the middlebox parses the DSCP field to determine if the packet has been sliced. It then parses the packet's token to obtain the index and the generation number of the corresponding payload stored in the array, which it verifies against the index and generation number in the \texttt{metadata\_table}. On successful verification, the middlebox retrieves the payload from the \texttt{payloads} array; re-attaches it to the packet; restores the \texttt{packet\_dscp}; clears the \texttt{ttl} field; and enqueues the packet for transmission to its destination.

\paragraph{Emulation platform performance scaling}
The emulation platform parallelizes the Slice \& Splice operations across multiple cores to keep up with the available network line rate.
We scale its processing by employing Receive Side Scaling (RSS).
We instantiate private instances of the aforementioned data structures per core to avoid synchronization overheads and employ symmetric RSS \cite{woo:scalable} to ensure that each packet is processed by the same core on its ingress and egress path.

\begin{lstlisting}[float,caption={Emulation platform's Slice \& Splice structures.},label={lst:emustructs},style=CStyle]
// struct to hold the payload
struct payload_t {
  uint16_t sz; //size of payload actually stored
  uint8_t packet_dscp;  //to restore on egress
  char blk[MAX_PAYLOAD]; 
};

// keep track of entry occupancy
struct metadata_t {
  uint64_t generation_number;
  uint8_t ttl; //ttl = 0 indicates an empty entry
};

// table of stored payloads
struct payload_t payloads[MAX_TABLE_SIZE];

// table of metadata structure
struct metadata_t metadata_table[MAX_TABLE_SIZE];

// metadata attached to the sliced packet
struct token_t {
  uint64_t key; //index and generation number
} token;

\end{lstlisting}

\section{Methodology}
\label{sec:method}

\paragraph{Experimental Platform}

\begin{figure}
    \centering
    \includegraphics[width=\linewidth]{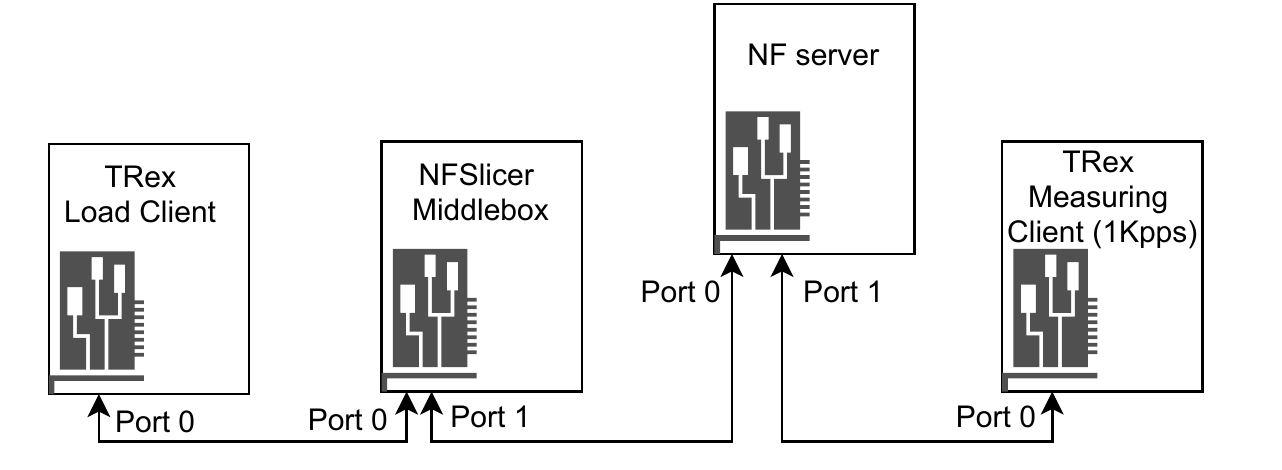}
    \caption{Experimental setup to measure end to end latency of a client in the presence of network load. }
    \label{fig:nfslicerBasicSetup}
\end{figure}

In order to isolate system effects as a consequence of packet sizes alone, we employ a  setup as illustrated in \cref{fig:nfslicerBasicSetup}. 
NFServer is the device under test, on which the evaluated NFs are deployed.
The \sys Middlebox emulates the Slice \& Splice functionality, as described in \cref{sec:impl}.
To collect end-to-end latency measurements that accurately represent server-side behavior, we employ two separate clients.
The Load Client offers knobs to configure the packet size and rate, thus controlling the NFServer's operational region/utilization.   
The Load Client's traffic flows to the NFServer through the Middlebox.

The Measuring Client's role is to take end-to-end latency measurements that accurately reflect the server-side latency effects on our envisioned production hardware-based \sys deployment, which is currently emulated in software by the Middlebox.
The Measuring Client is therefore deployed with the following two provisions.
First, to avoid measuring bias due to client-side queuing effects, the Measuring Client emits packets at a low fixed rate and measures each packet's end-to-end latency \cite{zhang:treadmill}.
Second, the Measuring Client's packets do not flow through the Middlebox, to avoid biasing measurements with the high latency of the software-based Slice \& Splice functionality's implementation.
Despite bypassing the Middlebox, our measured end-to-end latency is representative because a hardware-based \sys implementation would introduce a minuscule fixed latency overhead on each packet, as we show in \cref{sec:hardware-impl}.
Although the Measuring Client's packets do not get sliced, the vast majority of traffic received on the NFServer (\textgreater99.9\% in all our experiments) originates from the Load Client and \textit{does} get sliced. 
Therefore, this methodology allows us to accurately observe the systems-level effects of reduced on-server data movement, which is reflected on the resulting end-to-end latency of the Measuring Client's packets.
\textit{All latency numbers in the evaluation are reported from the Measuring Client,} averaged across three 30-second runs.

Both clients are based on the TRex scalable, open-source DPDK traffic generator \cite{trex}. Given our work's focus on large packets, the Measuring Client generates 1518B packets at 1Kpps, TRex's lowest available rate.
All four server-grade machines are 2-socket Intel Xeon 4214 processors with 256GB of RAM and a dual-port 100G ConnectX-6 Dx EN Mellanox NIC, running Ubuntu 16.04.7 LTS.
We disable all power management features and OS scheduling on the CPU cores involved in the experiments and only use cores and memory of the socket the NIC is directly attached to.

\begin{table}[b]

\begin{center}
\begin{footnotesize}
\begin{tabular}{ |p{1.6cm}|p{6cm}|}
\hline
 \textbf{NF} &\textbf{Description}\\
\hline 
 \multirow{2}{*}{L2 Forwarder} & Modifies ethernet addresses and is used as a bridge between interfaces \\
 \hline
 \multirow{2}{*}{QoS Metering} & Measures traffic arrival rate and classifies packets into groups of 
 corresponding rates \\
 \hline
 Firewall & Performs routing and access control \\
 \hline
 \multirow{2}{*}{VigNAT} & A formally verified Network Address Translator, which maps different IP address spaces\\
\hline
\end{tabular}
\caption{Shallow NFs used in our evaluation.}

\label{table:shallow-nf-desc}
\end{footnotesize}
\end{center}
\end{table}

\paragraph{Measurement tools}
In addition to TRex's in-built support for end-to-end latency measurements, we use \texttt{intel-cmt-cat} \cite{intel:cat} for microarchitectural event measurements, and Intel's Processor Counter Monitor \cite{pcm} toolset: \texttt{pcm-memory.x} for memory traffic and \texttt{pcm-pcie.x} for PCIe utilization. 
We also use \texttt{DDIOTune} \cite{fastclick:ddiotune} to configure DDIO in the microarchitectural study.

\begin{figure*}

	\includegraphics[width=0.75\linewidth]{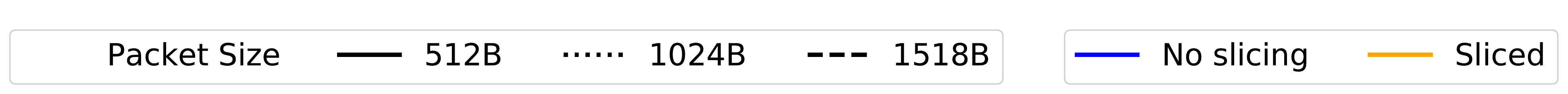}
	\vspace{-4mm}
	\centering
        
    \subfloat[L2 forwarder.]
    {
     \includegraphics[width=.185\linewidth]{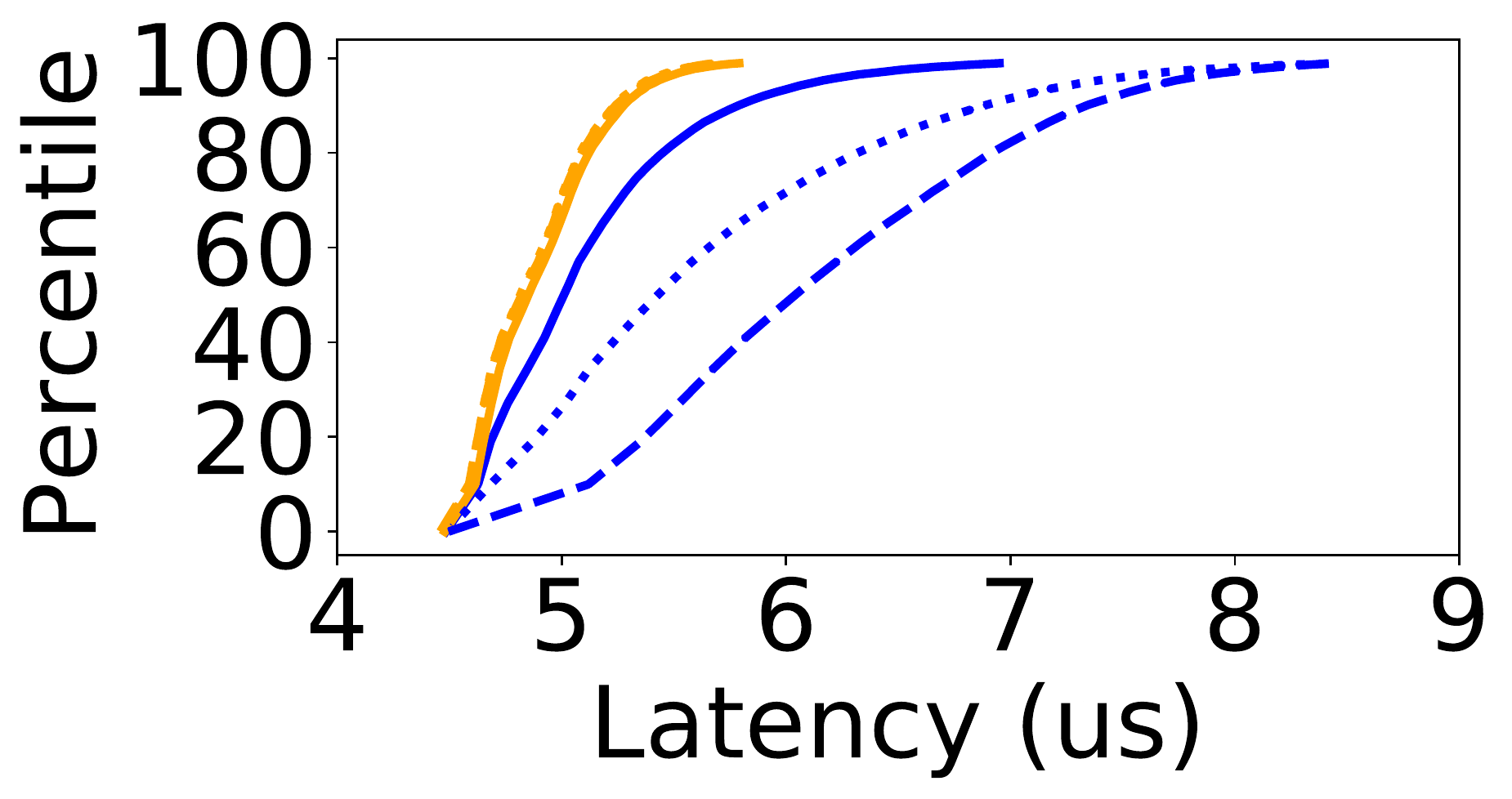}
	\label{fig:l2fwd-cdf}
    }
    \hfill
    \subfloat[QoS Metering.] {
		\includegraphics[width=.17\linewidth]{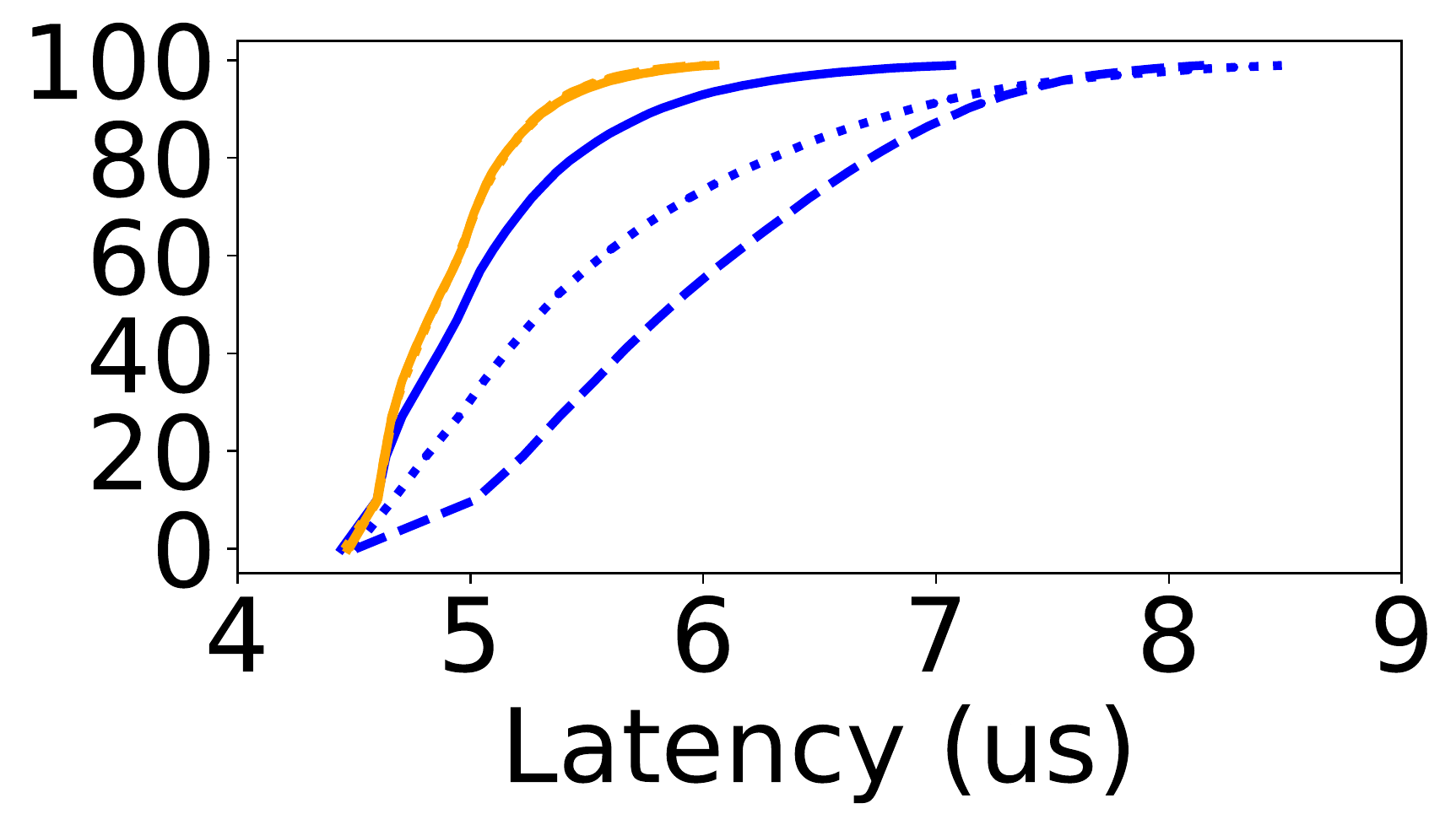}
		\label{fig:l3fwd-cdf}
	}
	\hfill
    \subfloat[Firewall.] {
		\includegraphics[width=.17\linewidth]{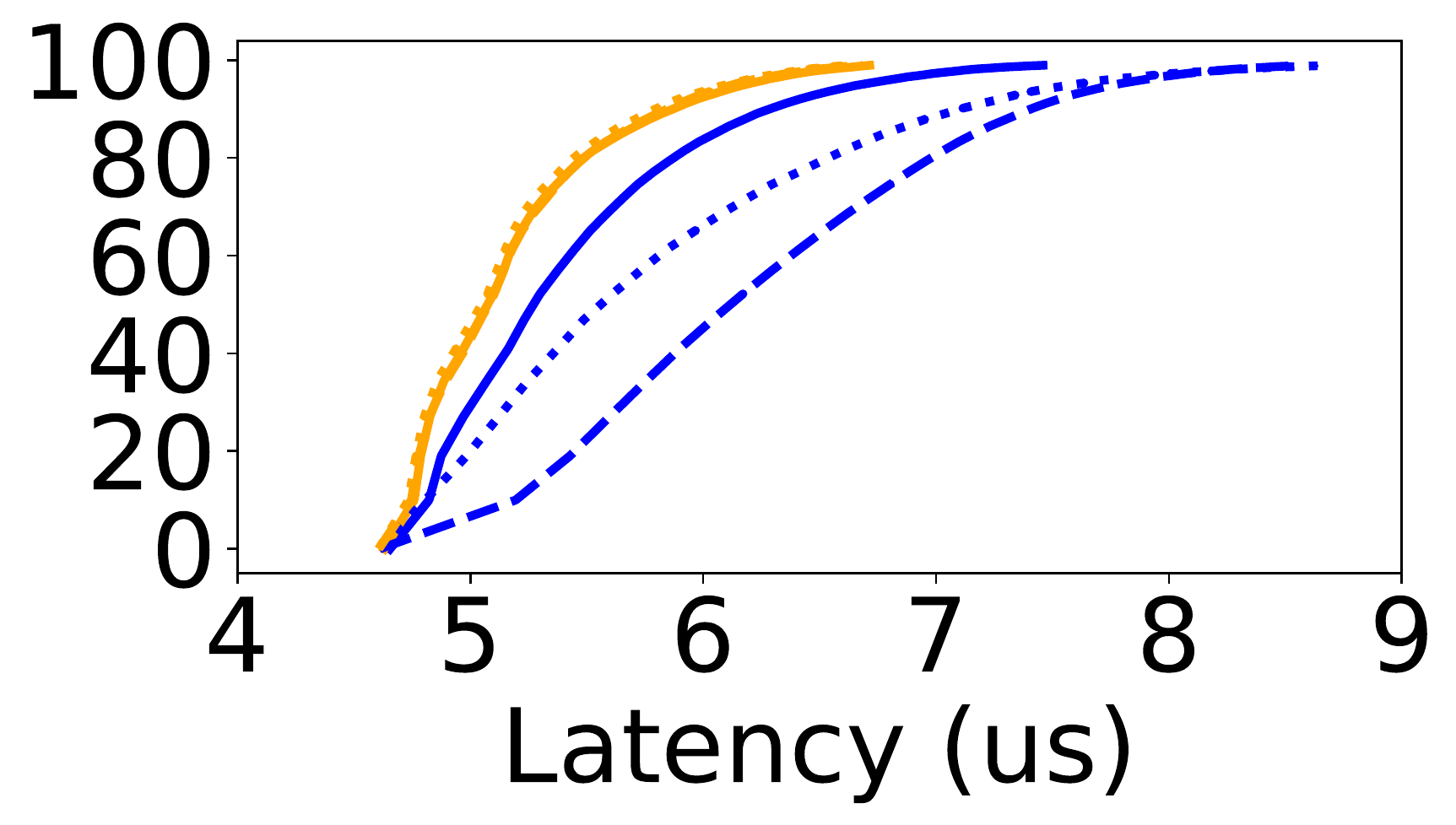}
		\label{fig:firewall-cdf}
	}
	\hfill
    \subfloat[VigNAT.] {
		\includegraphics[width=.17\linewidth]{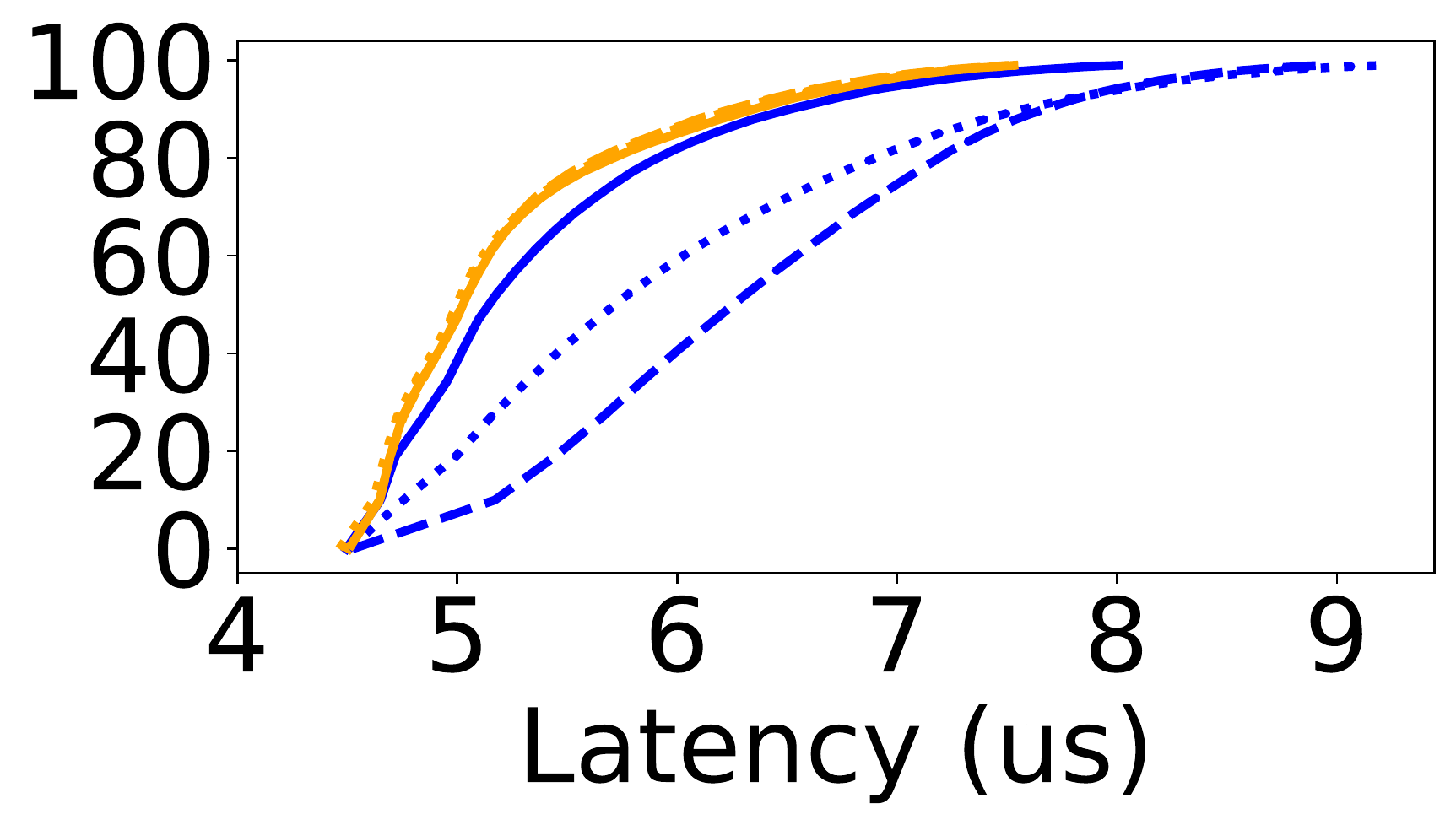}
		\label{fig:vignat-cdf}
	}
	\hfill
    \subfloat[NF chain.] {
		\includegraphics[width=.17\linewidth]{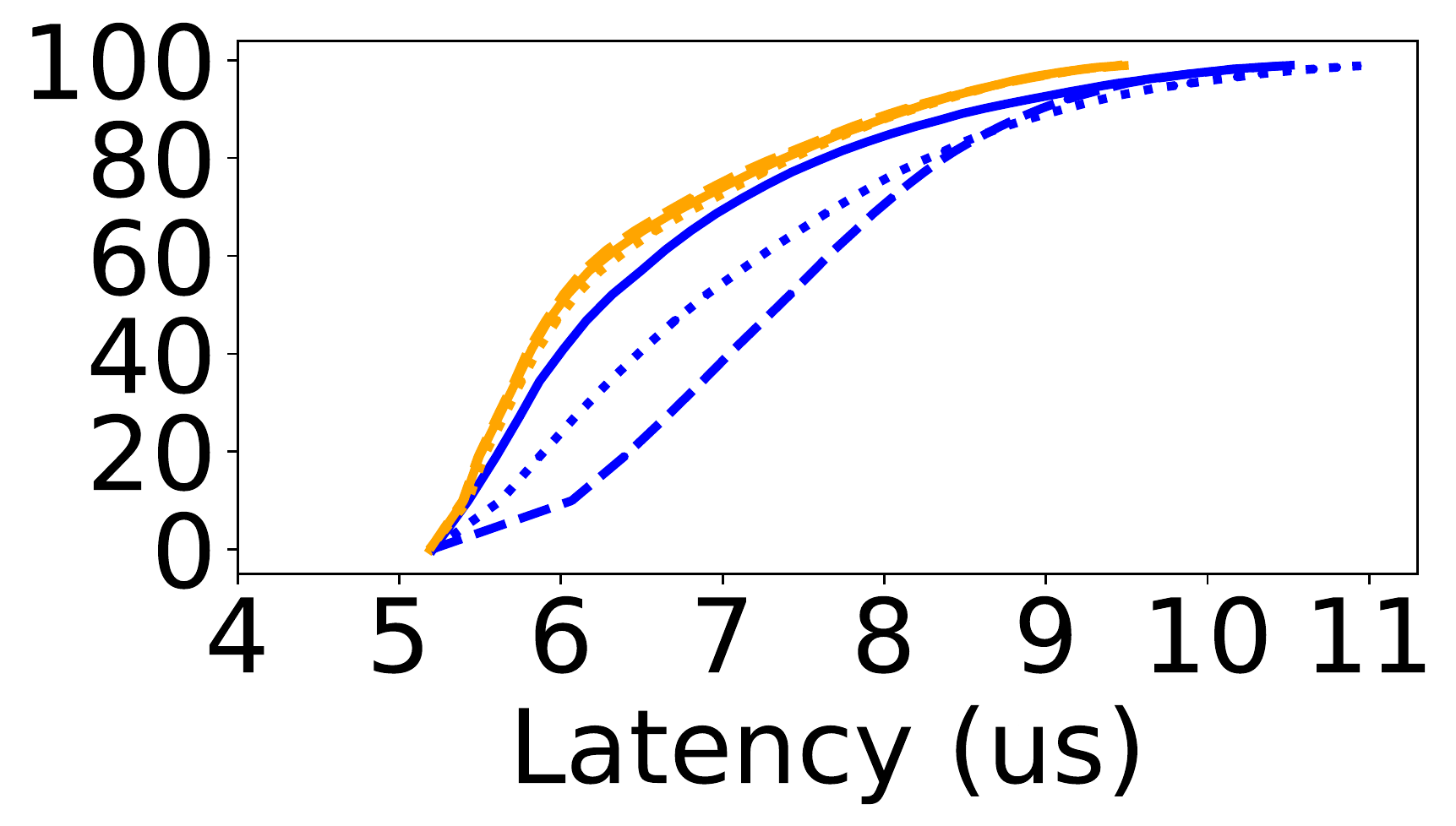}
		\label{fig:nfchain-cdf}
	}
	\caption{Latency CDFs for packets of different sizes processed by a range of shallow NFs at a fixed arrival rate of 4Mpps: baseline (\ie No slicing) versus sliced with \sys. The NF chain comprises Firewall followed by VigNAT.} 

	\label{fig:nf-cdfs}
\end{figure*}

\paragraph{Shallow NFs used} 
We evaluate four NFs which process L2--L4 headers.
In order to measure end-to-end packet latency, we configure the NFs to operate in loopback---i.e., regardless of the NF processing, the ingress and egress occurs through the server's same port.  \cref{table:shallow-nf-desc} briefly describes each NF's functionality. The L2 forwarder, QoS metering, and Firewall, are adapted from a set of applications provided by DPDK \cite{dpdk:sample-applications}. The fourth NF implements a formally verified NAT---VigNAT \cite{zaostrovnykh:vignat}. In addition to the four raw NFs, we also evaluate an NF service chain comprising  Firewall followed by VigNAT.
We developed the NFs and \sys emulation platform using DPDK version 20.02.1.

\section{Evaluation}
\label{sec:eval}

\begin{figure*}
	\vspace{-2mm}

    	\includegraphics[width=0.75\linewidth]{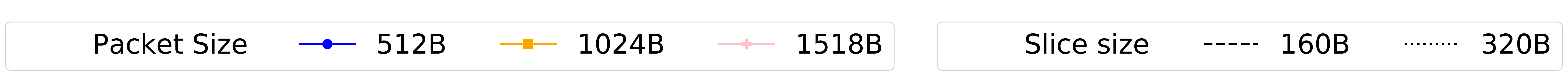}
	\vspace{-4mm}
	\centering
	
	\subfloat[L2 forwarder.]
    {
     \includegraphics[width=.20\linewidth]{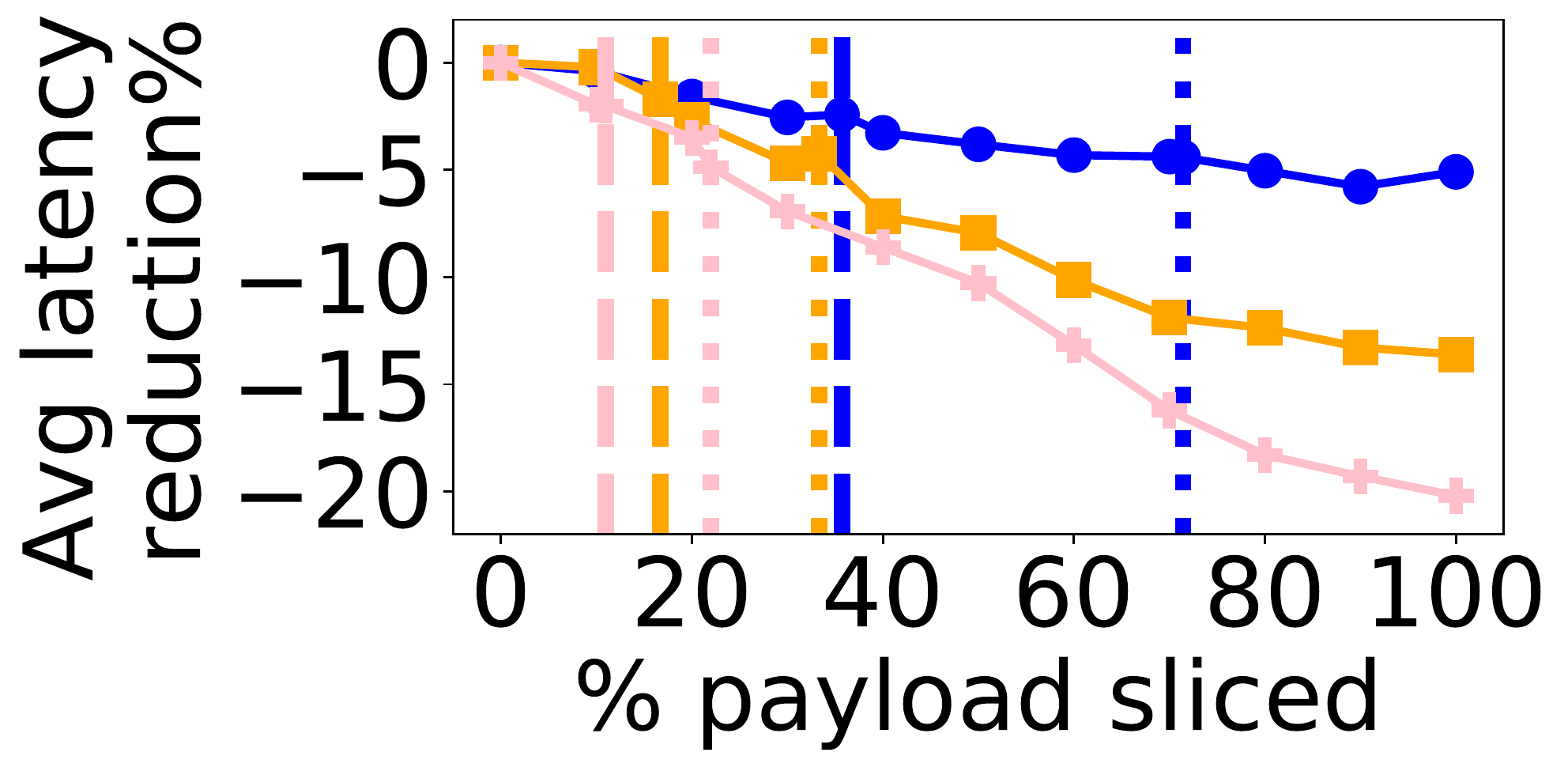}
     \label{fig:l2-forwarder-avg-lat}
    }
    \hfill
    \subfloat[QoS Metering.] {
		\includegraphics[width=.17\linewidth]{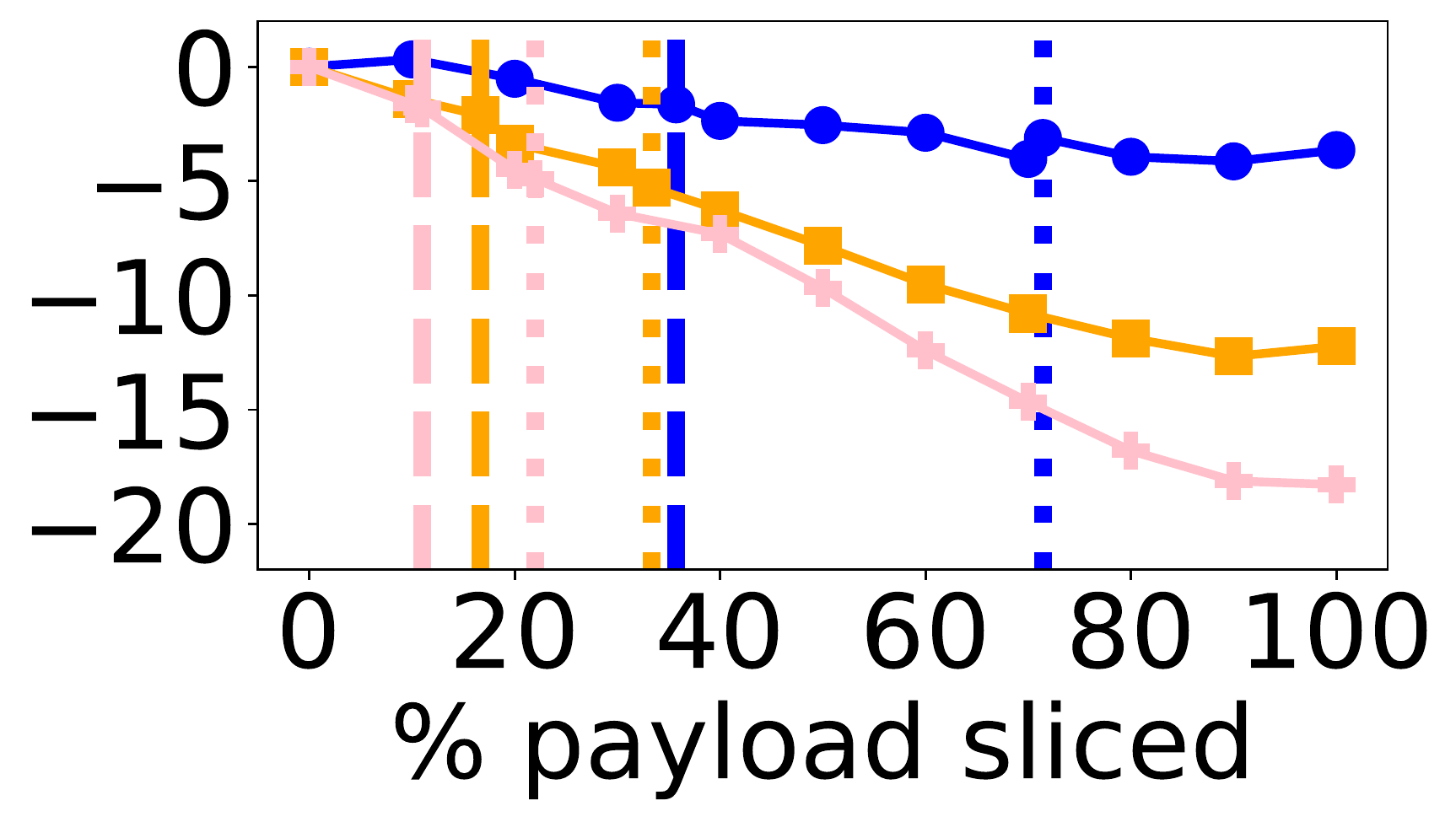}
	}
	\hfill
    \subfloat[Firewall.] {
		\includegraphics[width=.17\linewidth]{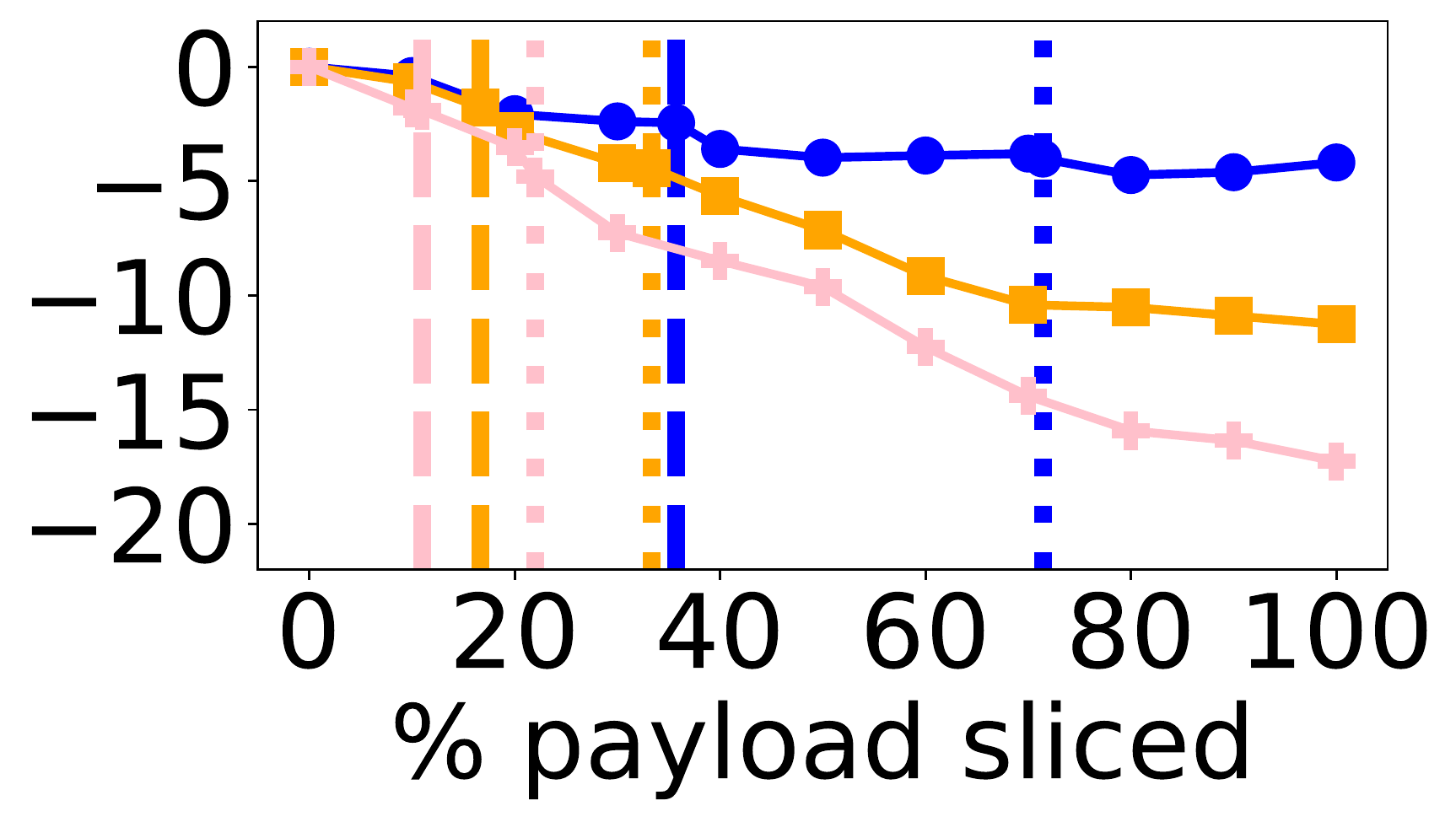}
	}
	\hfill
    \subfloat[VigNAT.] {
		\includegraphics[width=.17\linewidth]{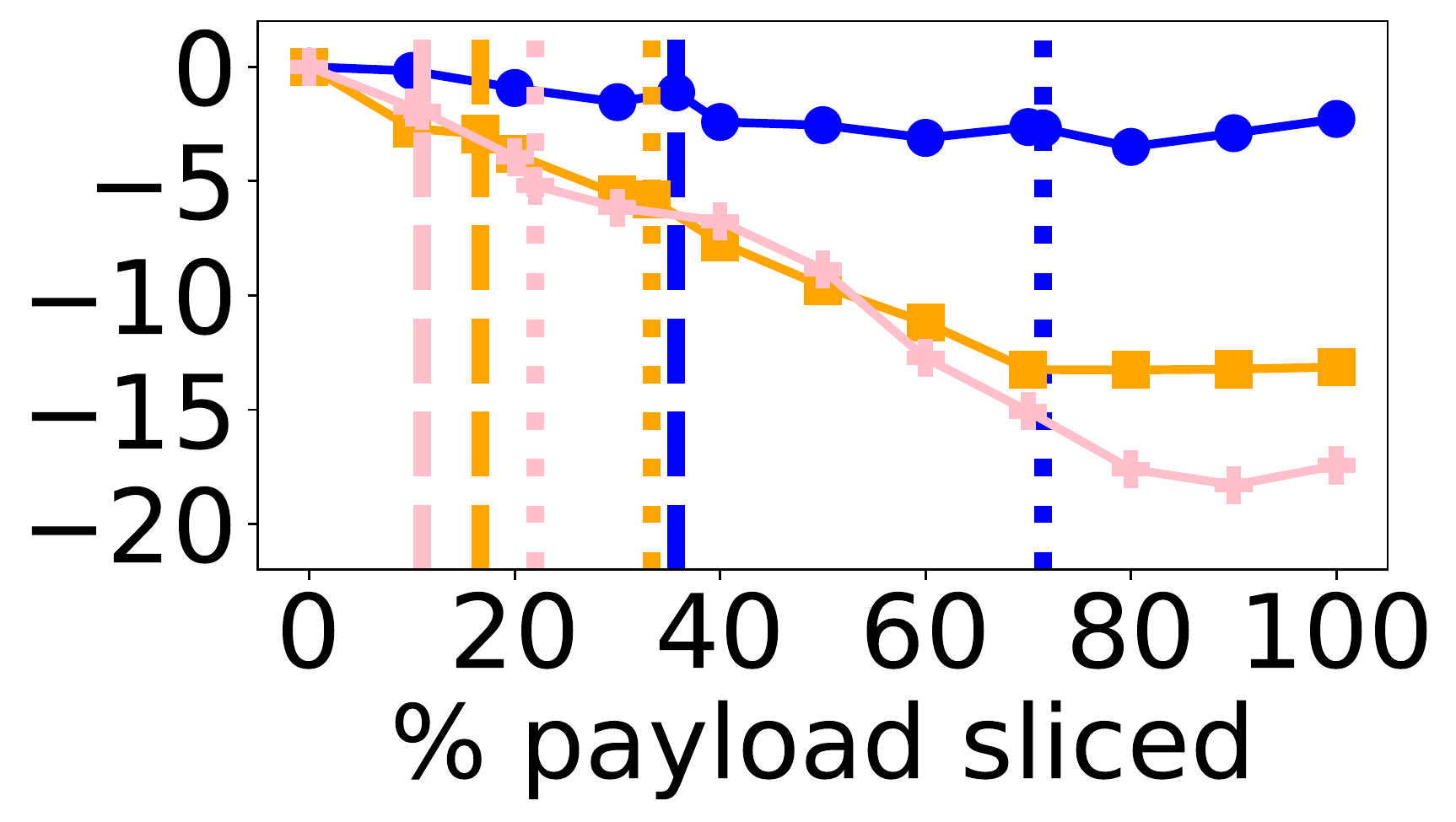}
	}
	\subfloat[NF chain.] {
		\includegraphics[width=.17\linewidth]{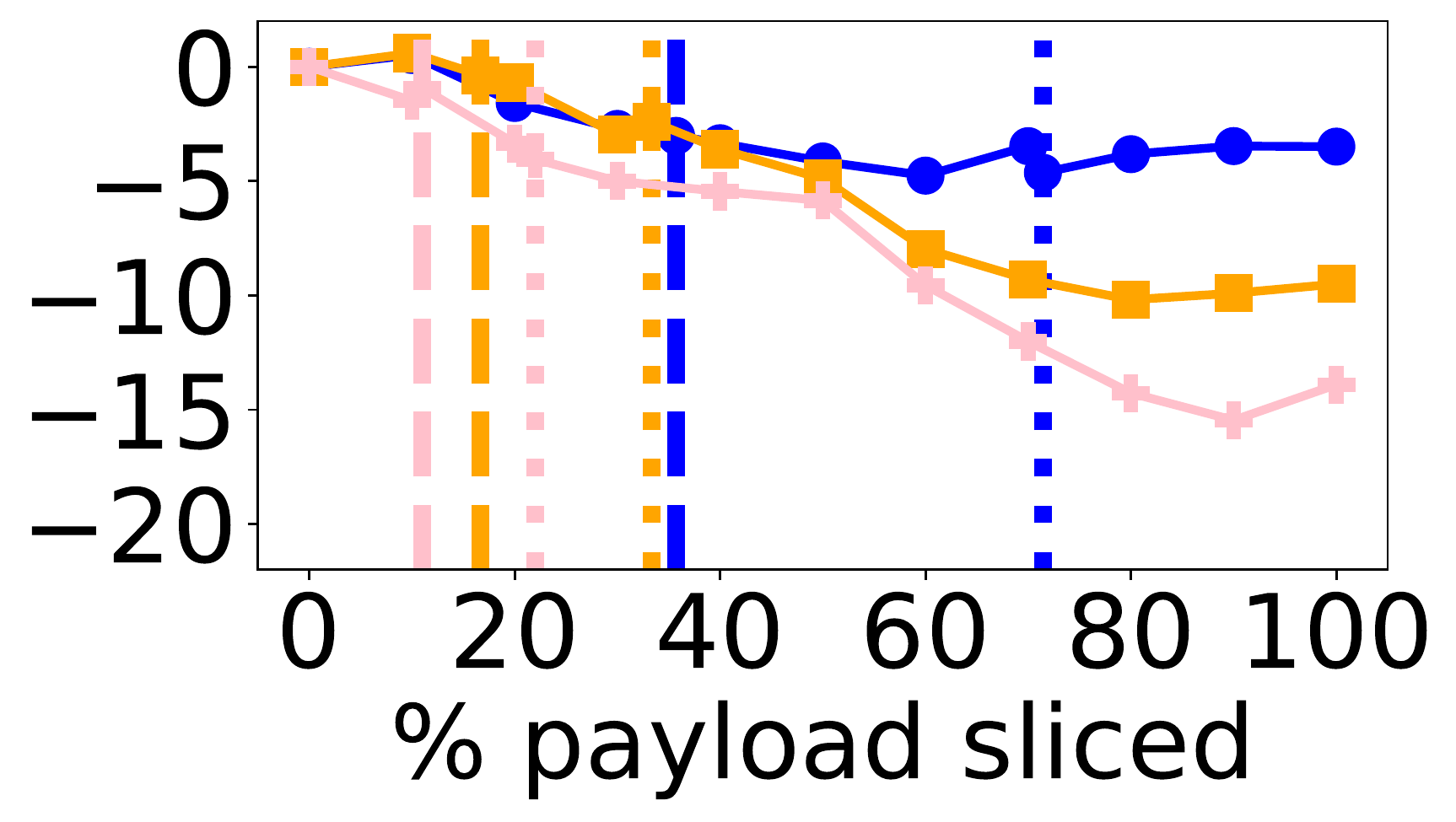}
	}

	\caption{Average latency reduction {compared to no slicing} as a function of fraction of payload sliced for shallow NFs at 4Mpps. Vertical lines mark the slicing fraction corresponding to 160B ({PayloadPark} \cite{goswami:parking}) and 320B of sliced payload per packet. 
}
\vspace{-2mm}
	\label{fig:latency-improvement}
\end{figure*}

This section first evaluates \sys's achieved latency reduction for a range of shallow NFs (\cref{sec:eval:nf-perf}).
We then demonstrate the sensitivity of latency improvement to the fraction of packet payload sliced (\cref{sec:eval:payload-size}) and packet arrival rate (\cref{sec:eval:packet-rate}), and compare \sys to a switch-based packet slicing approach like PayloadPark \cite{goswami:parking}.
Finally, we perform an extensive microarchitectural analysis to pinpoint the source of latency overhead for large packets, shedding light on the origins of \sys's achieved latency improvements (\cref{sec:eval:uarch}).

\subsection{Performance Impact of Packet Slicing}
\label{sec:eval:nf-perf}

\cref{fig:nf-cdfs} shows the response latency CDF for four different shallow NFs and an NF chain, ordered left to right by increasing computational intensity.
All plots are at 4Mpps (Million packets per second), the maximum packet arrival rate the emulation platform can sustain before its NIC saturates, as observed by non-negligible packet drops.
In all cases, the NFServer can sustain the packet rate with a single core. 

We observe similar trends across NFs. 
First, while the minimum latency for all packet sizes is the same ($\sim$4.5\microsecond), the larger the packets, the higher the NF's resulting response latency. 
Taking L2 forwarder as an example (\cref{fig:l2fwd-cdf}), 512B packets exhibit a p50 response latency of 5\microsecond, which grows to 5.4\microsecond for 1024B packets and to 6\microsecond for 1518B packets (see blue lines);
p90 response latency grows even faster with packet size, corresponding to 5.8\microsecond, {6.9\microsecond}, and {7.3\microsecond} for 512B, 1024B, and 1518B packets, respectively. 

Second and most importantly, by slicing packet payloads and preventing large network data transfers on and off the NFServer, \sys noticeably shifts the latency distribution of all packets to the left, \textit{equalizing all of them to the best-case latency of small (64B) packets, regardless of the original packet size} (see orange lines).

Third, the less computationally intensive the NF, the larger the latency gap between small and large packets, as processing events join data movement as a performance determinant.
To illustrate, the {3.8\% / 9.5\%} p50 / p90 latency gap between baseline and sliced handling of 512B packets in the case of L2 forwarder, shrinks to a {2.3\% / 4.1\%} p50 / p90 latency gap for the more computationally intensive VigNAT.
However, the latency gap for 1518B packets remains considerable regardless of the NF, ranging from {19.7\%} p50 and 18.5\% p90 for VigNAT to {20.2\% p50 and 28.6\%} p90 for L2 forwarder.
\sys also benefits our evaluated NF chain, which combines our two most compute-heavy NFs: Firewall and VigNAT.
\sys improves the 50p / 90p latency of 512B packets by 4.5\% / 5.6\% and of 1518B packets by 18.8\% / 9.4\%.
\textit{In summary, \sys delivers considerable end-to-end latency reduction for all shallow NFs, with benefits growing with packet size.}
Across all evaluated NFs, for 1518B packets, \sys improves 50p latency by 17.0--20.2\% and 90p latency by 9.4--28.6\%.

\subsection{Sensitivity to Payload Size Reduction and Comparison to Switch-Based Packet Slicing}
\label{sec:eval:payload-size}

\sys is designed to reduce the amount of data moved between the NIC and the server executing shallow NFs to the bare minimum.
A similar Slice \& Splice approach can also be implemented in programmable switches, as demonstrated in the recent work PayloadPark \cite{goswami:parking}. 
PayloadPark's intended goal is to boost the goodput of switch-server links, by slicing off a fixed-size chunk (160B) of each packet's payload.
A side-effect of the approach is that data movement on/off the server is also reduced, partially capturing \sys's benefits.
However, as we will demonstrate next, switch-based approaches face considerable limitations in terms of performance and scalability, and therefore cannot subsume \sys. 

\paragraph{Performance} We modify our emulation platform to slice a configurable fraction of each packet's payload, to measure the latency impact as a function of payload size reduction. 
\cref{fig:latency-improvement} shows the \textit{average} end-to-end latency reduction as a function of the amount of payload sliced, where 0\% represents the baseline case of full packet delivery (no slicing) and 100\% represents \sys's default operation, which slices each packet's whole payload, reducing all packets down to a fixed size of a single cache block (64B). 

While the general observations regarding \sys's opportunity as a function of packet size and for different NFs are similar to \cref{sec:eval:nf-perf}, \cref{fig:latency-improvement} additionally demonstrates that latency reduction is relative to the packet size reduction fraction---i.e., partial payload slicing will only yield a fraction of the benefits of full payload slicing.
Vertical dashed lines indicate the operational point of a switch-based solution like PayloadPark, which reduces packet payloads by 160B, corresponding to 36\% / 17\% / 11\% of 512B / 1024B / 1518B packets, respectively.
While slicing 160B captures most of the benefits for 512B packets, it misses out on significant opportunity for larger packets. 
For instance, for the L2 forwarder NF (\cref{fig:l2-forwarder-avg-lat}), slicing 11\% of a 1518B payload only reduces latency by {2\%}, instead of {17\%} with \sys's full payload slice.

PayloadPark's 160B slicing capability is partially imple\-men\-ta\-tion-specific and not a fundamental upper bound for every possible switch-based implementation.
However, switches have rigid performance constraints and the maximum amount of data they can slice without sacrificing throughput will always be bound by their SRAM interface's width and their provisioned RMT pipeline's number of stages.
To illustrate, a hypothetical switch-based implementation with double the payload slicing capability (320B instead of 160B, marked by {vertical} dotted lines in \cref{fig:latency-improvement}) roughly doubles the latency improvement potential of switch-based solutions, but still leaves \sys with a considerable headway.
Using the same L2 forwarder example, doubling slicing capability to 320B improves latency reduction for 1518B packets to {4.8\%}, still leaving a {$3.6\times$} gap with \sys.
Furthermore, while \cref{fig:latency-improvement} shows \textit{average} latency, the improvement opportunity is even larger for tail latency; for instance, \sys's p90 latency improvement for 1518B packets is {$5.6\times$} higher compared to a hypothetical switch-based implementation slicing 320B per packet (not shown due to space constraints).

\begin{figure}
    \centering
    \includegraphics[width=0.8\linewidth]{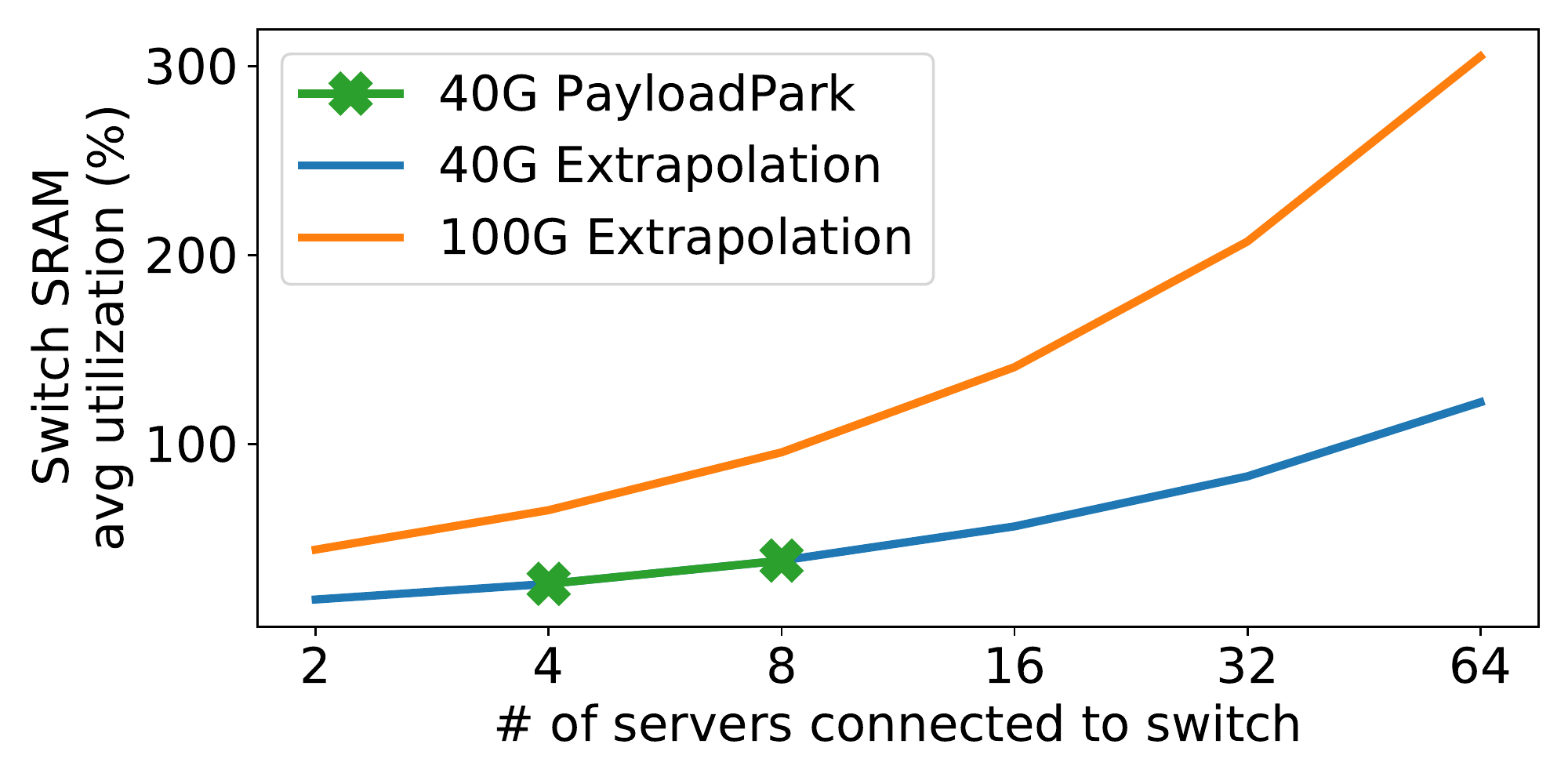}
    \caption{Scalability study of SRAM requirements for a switch-based implementation, assuming a 100G switch.  Blue/orange lines correspond to servers with 40G/100G NICs, respectively.}
    \label{fig:sram-util}
\end{figure}

\paragraph{Scalability}
In addition to the aforementioned performance limitations, switch-based approaches have inherent scalability limitations compared to \sys. 
While a switch-based solution's hardware requirements grow  with  the cluster size, \sys's scalability is virtually infinite, as hardware requirements on each server's NIC are unaffected by deployment scale.
\cref{fig:sram-util} demonstrates this fact with a first-order model, instrumented with data derived from PayloadPark~\cite{goswami:parking-arxiv}, which offers two data points: the \textit{average} SRAM utilization of the used 100G switch is 26\% and 38\% with 4 and 8 40G NIC NF servers, respectively.
The most {optimistic} extrapolation indicates that switch-based slicing can support up to 38 40G servers.
Furthermore, a simple application of Little's Law suggests that by upgrading the NICs on the servers from 40G to 100G (to mirror our experimental evaluation setup), the switch can only support up to 8 servers.

In conclusion, although switch-based approaches like PayloadPark and NIC-based approaches like \sys share mechanics, their effects on system behavior optimization only partially overlap. \sys is capable of slicing a packet's whole payload---rather than only a small subset---at line rate, thus \textit{completely ameliorating detrimental performance effects due to excess data movement}. \sys also \textit{scales} perfectly with cluster size, but, unlike PayloadPark, does not improve link {goodput}, as the entire packet traverses the link between the switch and the server.
Given the complementary strengths of switch-based and NIC-based approaches, they can be combined to achieve all three desirable qualities: data movement minimization, link goodput improvement, and scalability.

\subsection{Sensitivity to Packet Arrival Rate}
\label{sec:eval:packet-rate}

The latency gap between small and large packet processing grows with the packet arrival rate. 
Unfortunately, the increased network bandwidth requirements on our emulation platform's Middlebox (which needs to sustain $2\times$ the target packet arrival/transmission rate) limit the peak sustainable arrival rate of 1518B packets to 4Mpps.
Therefore, we demonstrate the effect of this growing gap as a function of packet arrival rate by removing the \sys Middlebox and directly connecting the Load Client to the NF server, which allows us to almost double the peak packet rate to 7Mpps.

\begin{figure}
    \centering
    \includegraphics[width=.9\linewidth]{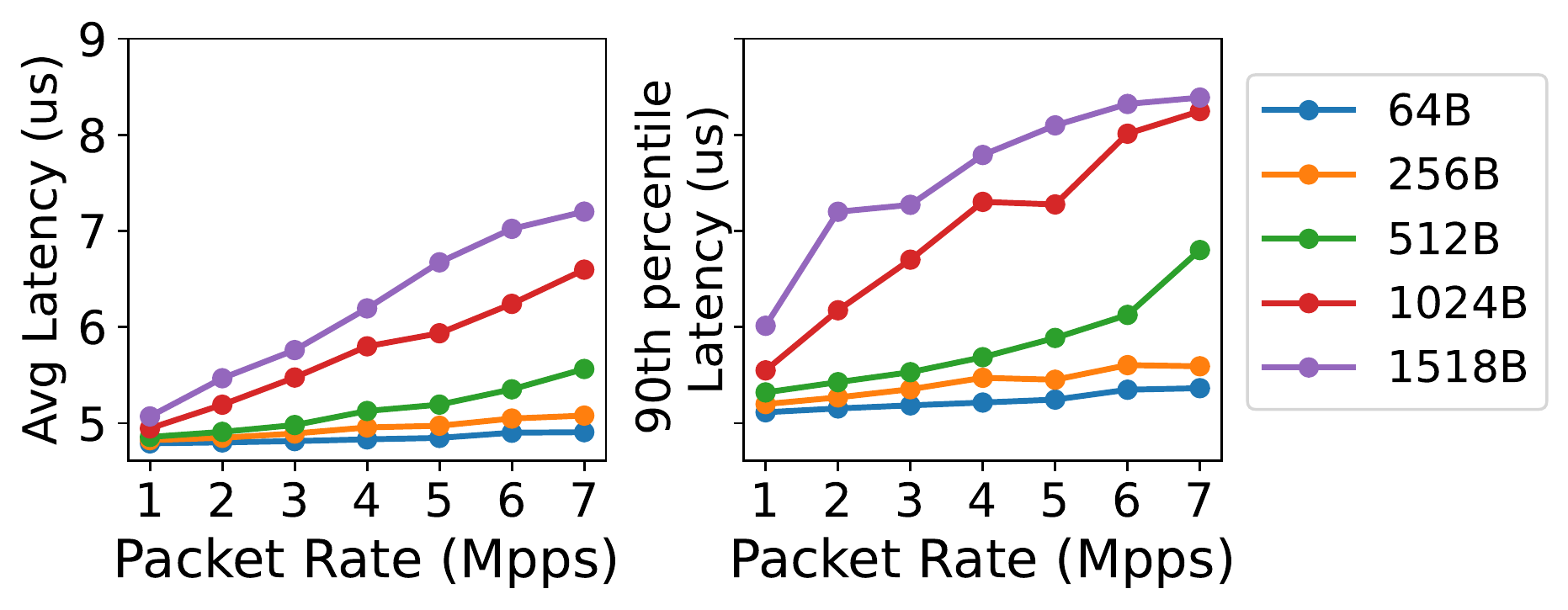}
    \caption{Latency gap between small and large packets growing with packet arrival rate. Results for an L2 forwarder NF. }
    \label{fig:loadlatency}
\end{figure}

\cref{fig:loadlatency} shows the average and p90 latency as a function of packet rate for an L2 forwarder NF. 
The $1.2\times$ average latency gap between 64B and 1518B packets at 4Mpps is reduced to $1.05\times$ at 1Mpps, but grows to $1.46\times$ at 7Mpps.
A similar, and more pronounced, trend appears for p90 latency:
the $1.4\times$ gap between 64B and 1518B packets at 4Mpps drops to $1.08\times$ at 1Mpps, but grows to $1.55\times$ at 7Mpps.

In addition, we experimentally verify that for the packet rate range sustainable by the emulation platform (1--4Mpps), latency results of a 1518B-packet stream that reaches the NFserver after getting sliced by the middlebox are equivalent to those of a direct client-to-server 64B packet stream.
In other words, from the measuring client's perspective, \textit{loading the NF server with large packets that are sliced by \sys down to 64B is equivalent to directly loading the server with 64B packets, without any prior slicing involved.}
We use this equivalence throughout the {next} section of our evaluation to study on-server effects of full-size versus sliced large packets at a fixed packet rate of 6.5Mpps, which is significantly higher than the emulation platform's peak sustainable rate of 4Mpps.

\begin{figure}[t]
    \centering
    \includegraphics[width=.65\linewidth]{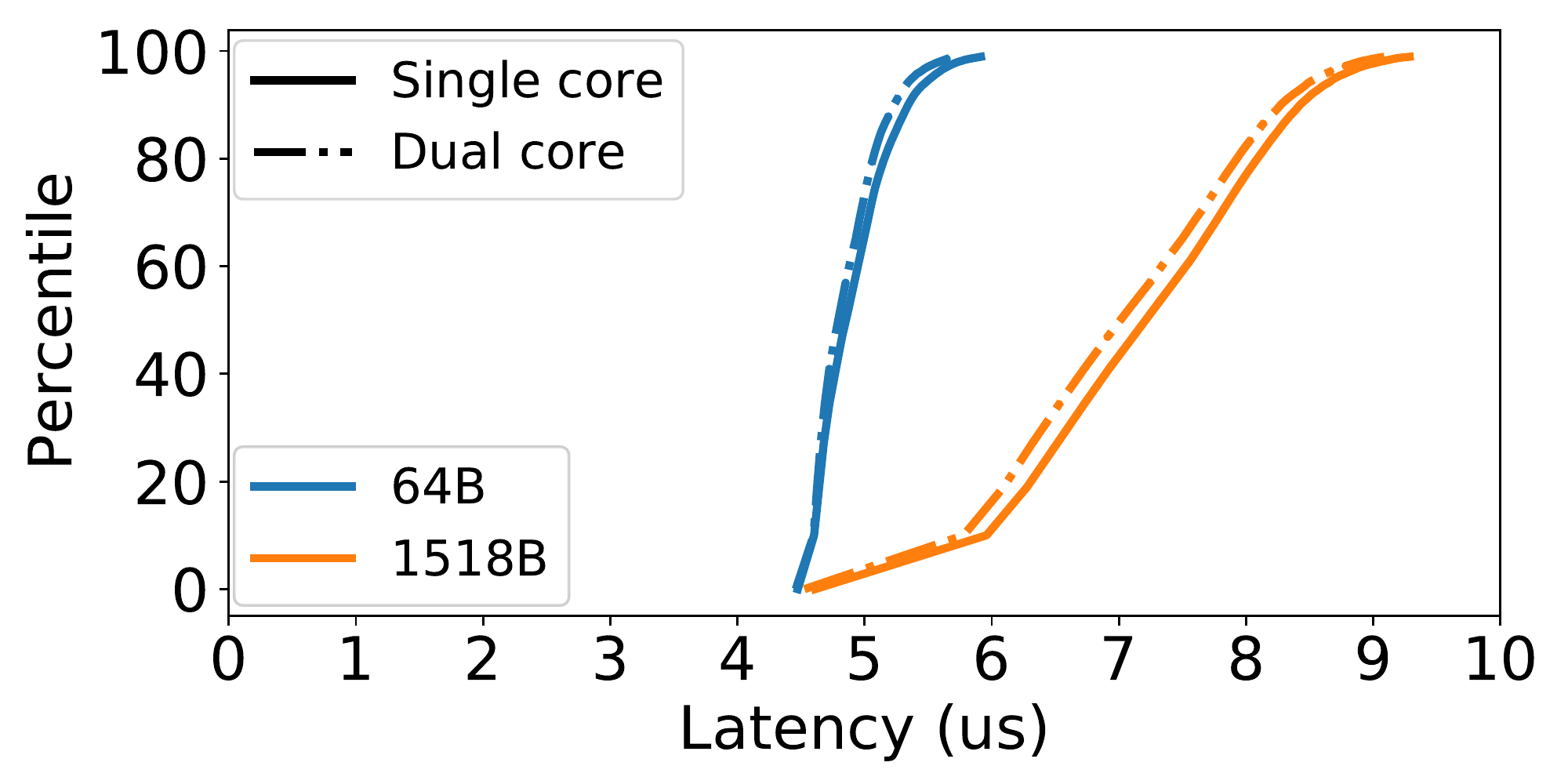}
    \caption{L2 forwarder latency CDF for 64B and 1518B packets at 6.5Mpps. Using a dedicated core to handle the measuring client's traffic does not affect the latency gap between small and large packets.
}
    \label{fig:dualCoresBaselineExp}
\end{figure}

\subsection{Microarchitectural Study}
\label{sec:eval:uarch}

We now embark to pinpoint the underlying sources of the performance gap between small and large packet handling.
Given the high similarity in behavior across the evaluated NFs, we focus on L2 forwarder as a representative NF {for this in-depth microarchitectural study.} 
Considering the microarchitectural components exercised by the path of packets processed by an NF, the culprit may be any of the following: the CPU, the LLC, 
the memory, and the NIC-processor I/O interface (i.e., PCIe).
We perform a series of experiments to isolate the impact of each component.

\begin{figure}[t]
    \centering
    \includegraphics[width=.7\linewidth]{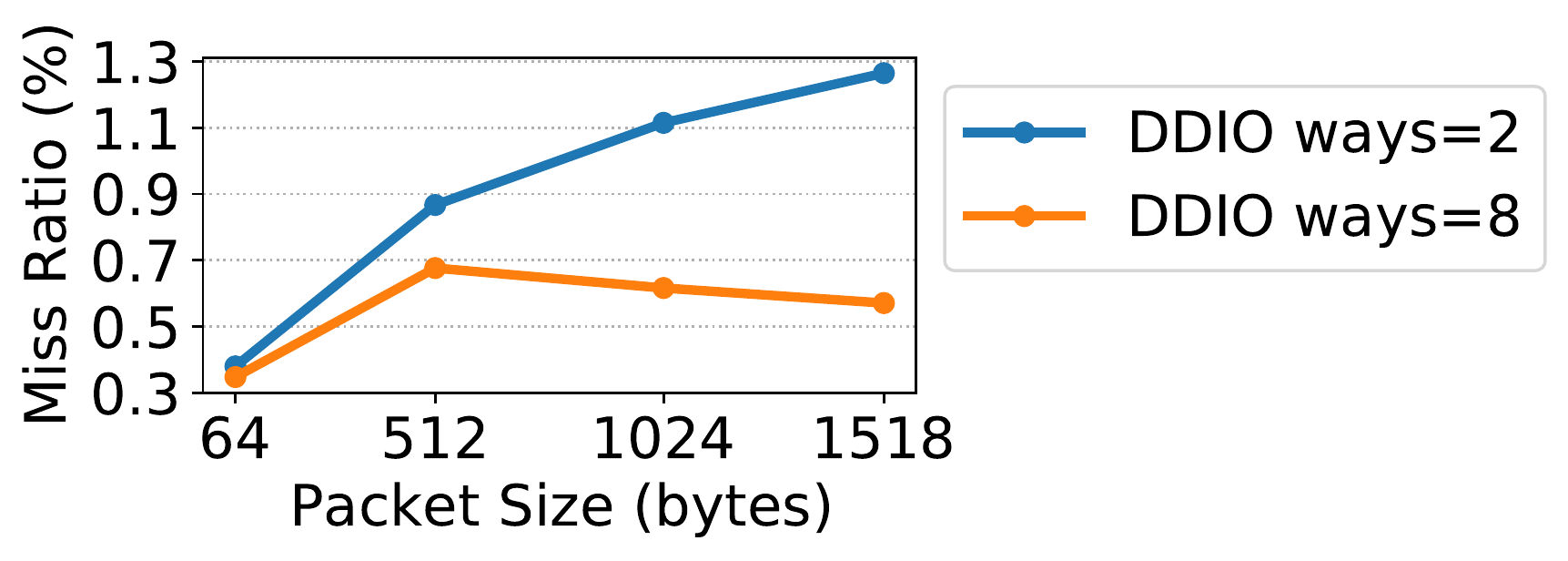}
    \caption{LLC miss ratios for varying packet sizes.}
    \label{fig:llcMissRatio}
\end{figure}
\begin{figure}[t]
    \centering
    \includegraphics[width=.65\linewidth]{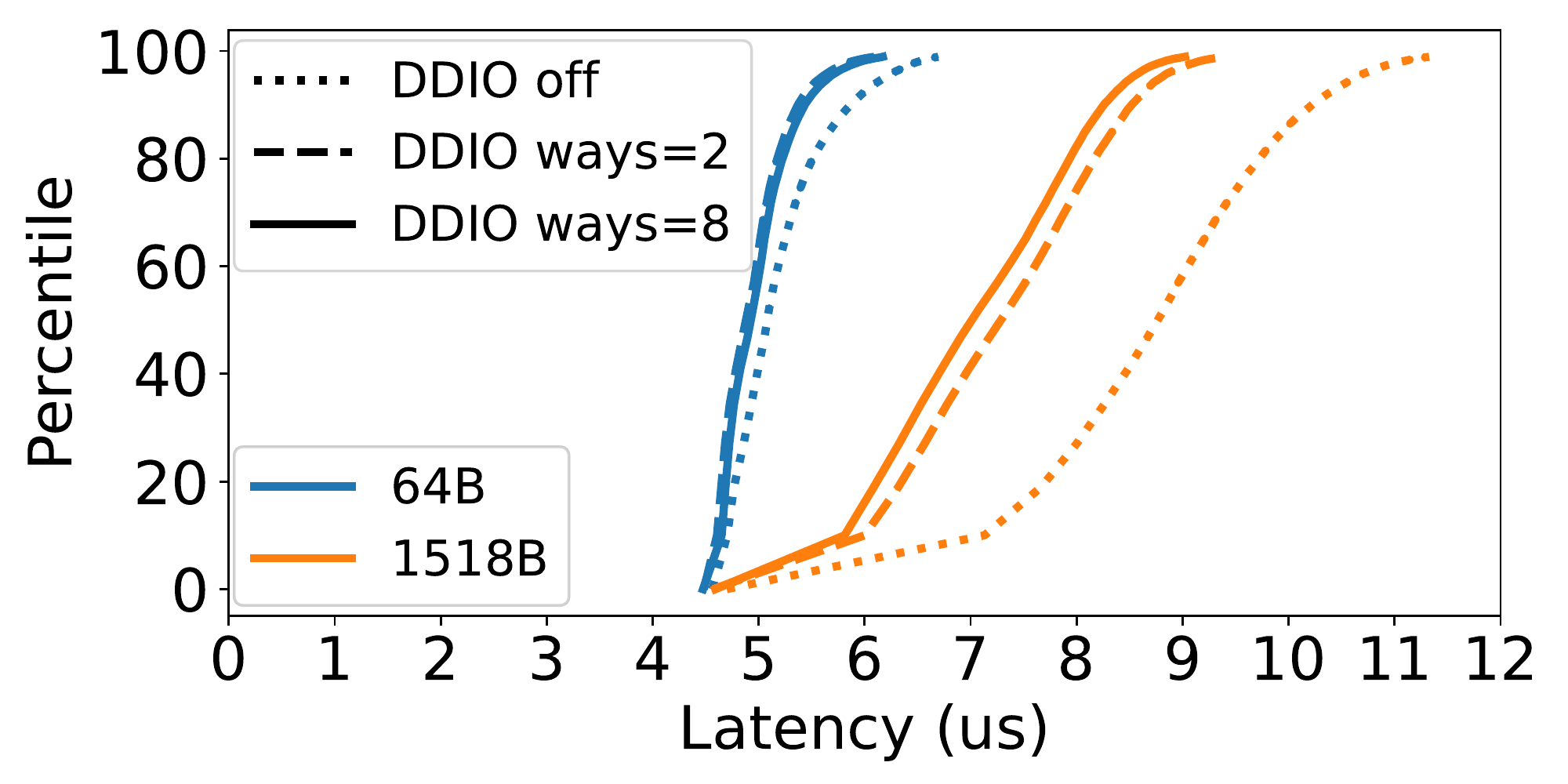}
    \caption{L2 forwarder latency CDF of 64B and 1518B packet under different DDIO configurations.}
    \label{fig:latencyDDIOoff}
\end{figure}

\paragraph{CPU} Although per-packet processing requirements for our NFs of focus are insensitive to packet size, we investigate whether larger packet size introduces adverse queuing effects at the core or private caches (L1/L2), by separating the load and measuring streams to be handled on different cores (both on the socket local to the NIC). 
\cref{fig:dualCoresBaselineExp} shows the results.
Unsurprisingly, both latency curves slightly shift to the left, as the measuring client now receives responses from a dedicated core.
However, the latency improvement with the addition of a second core is minimal and, most importantly, the relative latency gap between small and large packets remains unchanged.
We thus conclude that the performance gap between small and large packets \textit{is not attributed to a processing bottleneck or contention in private caches.}

\paragraph{Cache (LLC)}
Modern DDIO technology \cite{ddio} steers incoming packets directly into a portion of the LLC.
In cases of extreme contention, incoming packets may be evicted from the LLC to memory before they are consumed by a core. This effect, known as ``leaky DMA'' \cite{tootoonchian:resq} can have adverse effect on performance. 
We investigate if that's the case with growing packet size, by studying LLC behavior.

\cref{fig:llcMissRatio} shows the L2 forwarder NF's LLC miss ratio under two DDIO configurations: (i) the default one, which dedicates two LLC ways for network traffic injection\footnote{Unless stated otherwise, this default DDIO configuration has been used throughout this paper's evaluations.}, and (ii) an 8-way DDIO configuration.
We observe that a 20-fold increase in packet size results in a $3\times$ miss ratio increase under the default 2-way DDIO configuration.
However, the absolute LLC miss ratio remains negligible (less than $1\%$) in all cases, therefore, it cannot be a performance
determinant responsible for a 55\% gap in end-to-end p90 latency (see dashed lines in \cref{fig:latencyDDIOoff}).
This assessment is further confirmed by the 8-way DDIO configuration: the LLC miss ratio for large packets is $2.2\times$ lower than 2-way DDIO (\cref{fig:llcMissRatio}), but the small vs. large packet latency gap is largely insensitive to the DDIO configuration (dashed vs. solid lines in \cref{fig:latencyDDIOoff}). 
We thus conclude that {larger packets do not result in noteworthy LLC contention}, so \textit{LLC behavior is not the latency gap's culprit.}

\begin{figure}[t]
    \centering
    \includegraphics[width=.7\linewidth]{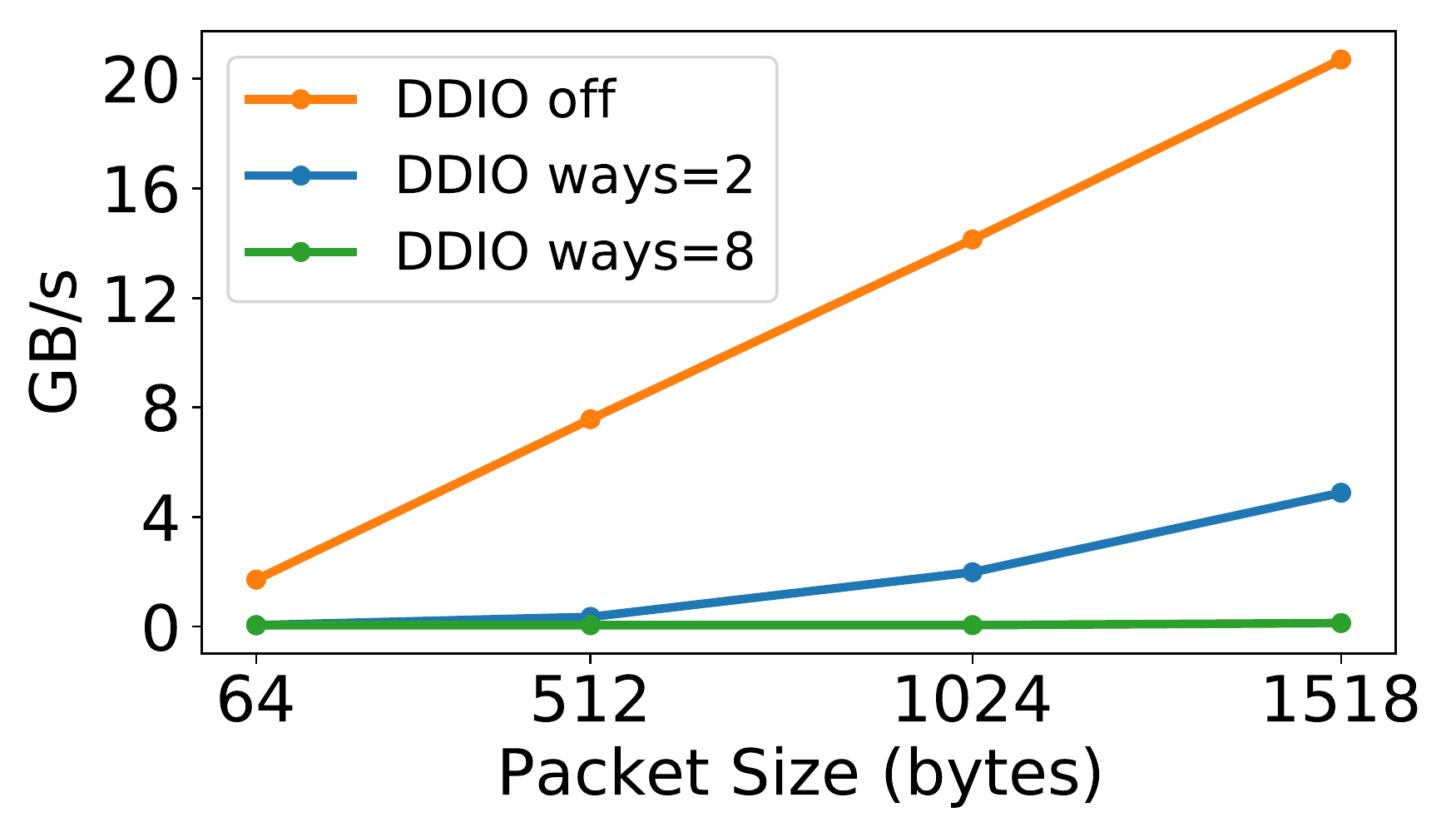}
    \caption{Total memory bandwidth utilization under different DDIO configurations.}
    \vspace{-2mm}
    \label{fig:socketMemBw}
\end{figure}

\paragraph{Memory bandwidth utilization}

\cref{fig:socketMemBw} displays the memory bandwidth consumed under three different DDIO configurations: 2-way (default), 8-way, and off.
The 2-way DDIO configuration offers enough cache capacity to keep all packets up to 512B LLC-resident, \iscanew{resulting in no memory traffic}. For larger packets, some network data spills to memory, generating non-zero bandwidth consumption, which, however, is too modest to justify the latency gap between small and large packets.
In contrast to 2-way DDIO, \cref{fig:socketMemBw}  shows that 8-way DDIO completely captures network data movement within the LLC, as memory bandwidth use remains zero for all packet sizes.
Despite that difference, \cref{fig:latencyDDIOoff} shows that the 1518B-packet latency curves almost overlap for 2-way and 8-way DDIO.
Hence, the latency gap between small and large packet handling \textit{is not attributed to memory effects.}

We also study the \textit{DDIO off} configuration as an interesting case where memory bandwidth usage \textit{does} affect the latency gap between small and large packets. 
When DDIO is disabled, all network traffic moves on and off the server through memory, resulting in significant memory bandwidth usage, as shown in \cref{fig:socketMemBw}.
\cref{fig:latencyDDIOoff} shows that putting memory accesses on the critical path has a direct impact on latency, even more so for larger packets, where the increased latency effect is amplified due to increased queuing on highly contended memory. The \textit{additional} increase in the latency gap between small and large packets in the \textit{DDIO off} configuration (dotted lines in \cref{fig:latencyDDIOoff}) is attributed to memory bandwidth contention.

\begin{figure}[t]
    \centering
    \includegraphics[width=0.6\linewidth]{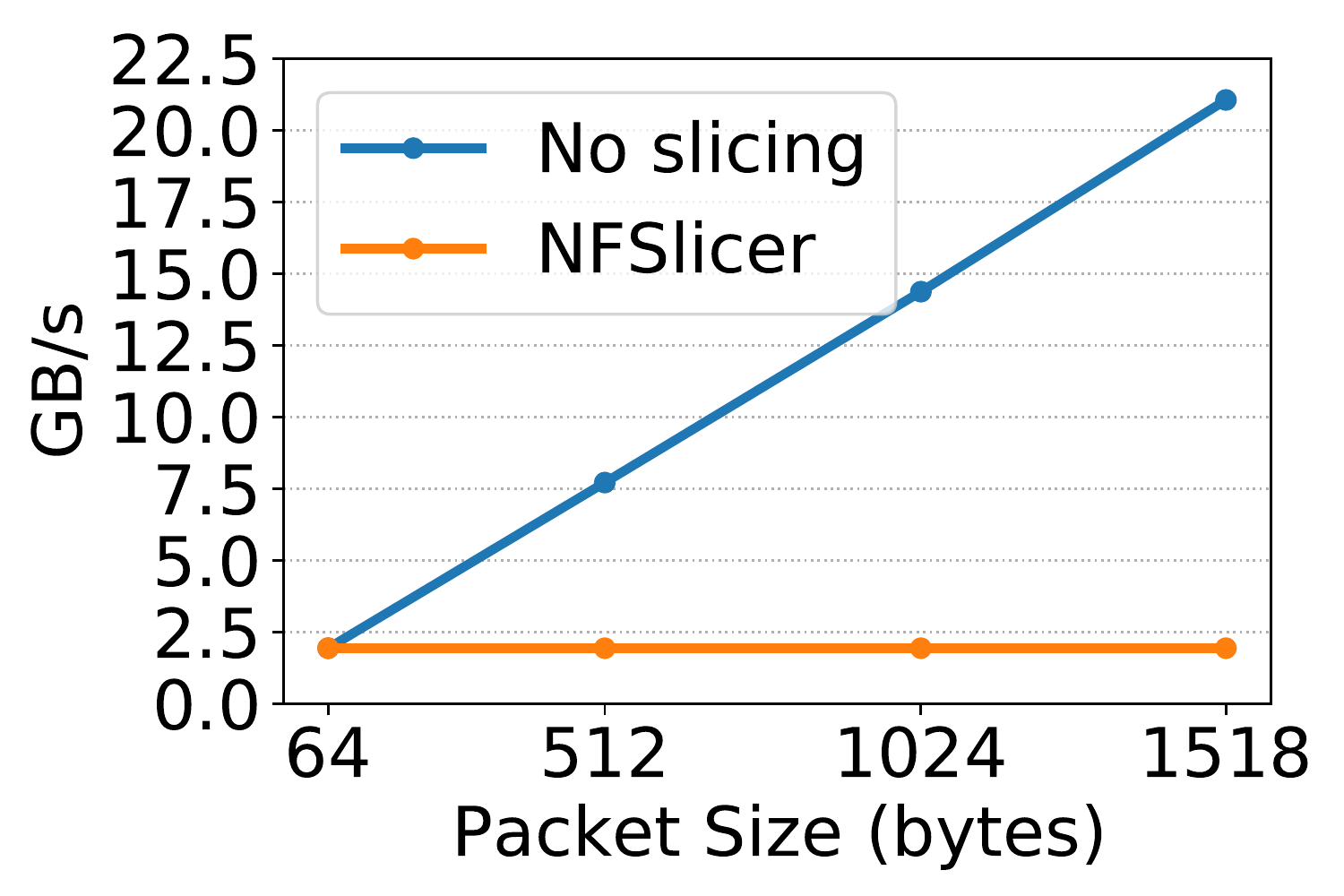}
    \caption{NIC-generated DMA traffic over PCIe.}
    \label{fig:pcieReadsWrites}
\end{figure}

\paragraph{PCIe utilization}
\cref{fig:pcieReadsWrites} shows PCIe utilization as a function of packet size.
While, as expected, PCIe utilization grows linearly with packet size, this trend clearly contrasts trends of memory and cache miss ratio behavior (when DDIO is enabled). 
The NIC moves entire packets over PCIe to the on-server memory hierarchy, regardless of the fact that the NF executing on the CPU only operates on a small fraction of the entire packet.
Aggregate data movement between the NIC and server memory hierarchy grows to 21GBps, or 67\% 
of the theoretical peak of the PCIe3x16 interface our NIC is attached to.
At such high utilization, it is common for severe queuing effects to emerge, directly hurting the latency of individual packets.
In contrast, \sys results in a constant low PCIe bandwidth use of only 1.9GBps, regardless of packet size.

Given our previous microarchitectural analyses showing that the processor, private caches,  LLC, and memory are far from being performance bottlenecks, \textit{we conclude that the most likely culprit for the significant latency gap between small and large packets is data movement over PCIe}. 
\sys's Slice \& Splice approach effectively alleviates this data movement bottleneck. By selectively transferring only the fraction of the packet that is needed by the deployed NF, it reduces data movement over PCIe by up to $24\times$, bridging the latency gap between small and large packet handling.

\section{Toward a Hardware \sys Implementation}
\label{sec:hardware-impl}

\begin{figure*}[t]
    \centering
    \includegraphics[width=.9\textwidth]{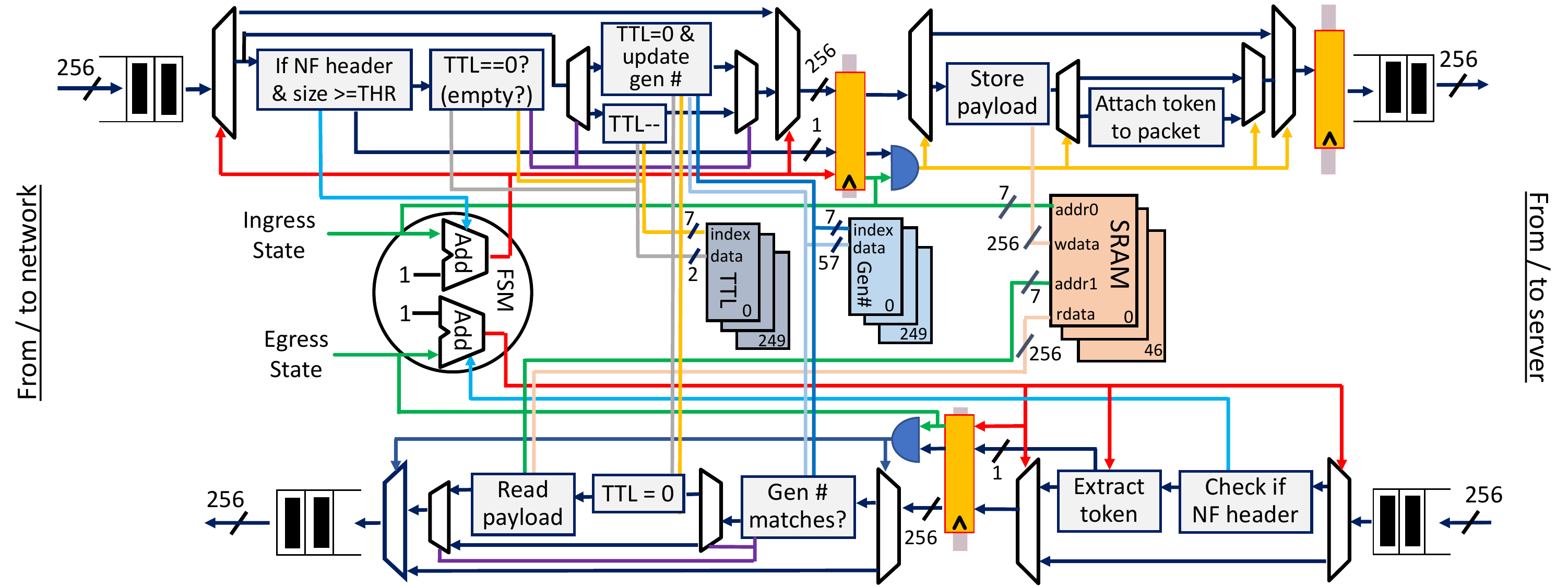}
    \caption{Architectural diagram of ingress (upper half) and egress (lower half) Slice \& Splice pipelines.} 
    \label{fig:pipeline}
\end{figure*}

{
We envision \sys as a hardware IP block on future NICs. 
Though the focus of this paper has been to present a thorough study of the system-level effects due to high network data transfer rates and the demonstration of \sys's performance improvement promise, we have also embarked on preliminary hardware design and synthesis to showcase \sys's feasibility for hardware implementation, and to provide area and power estimates.

\sys's pipeline consists of simple logic; the main hardware cost comes in the form of payload storage, which is implemented as SRAM. 
In this section, we describe a hardware implementation for a 100G NIC and use Synopsys tools to acquire area, power, and timing estimates.
We then outline strategies for scaling this IP block up for higher line rates.

We provision the SRAM as a static array of $N$ $M$-byte entries, sized to accommodate the bandwidth-delay product of our target NF deployments.
$N$ is dictated by the maximum packet arrival rate. As \sys only slices packets $\geq$ 500B, the maximum packet arrival rate at 100G is $\frac{1}{40ns}$.
$M$ is dictated by the largest packet payload \sys needs to accommodate, corresponding to 1518B packets.
Finally, the highest average end-to-end latency in our experiments is 9.3\microsecond for the NF chain, which also includes the roundtrip latency between the server and the client. 
By over-provisioning for an average service time of 10\microsecond, $N=\frac{10\microsecond}{40ns}=250$.
Overall, our hardware design provisions $N \times M~bytes = 355KB$ of SRAM for packet payload storage, which---to put things into perspective---is smaller than a private L2 cache of a modern multicore server.

\cref{fig:pipeline} shows the high-level architecture of the ingress and egress pipelines of our \sys hardware design. 
After verifying our design's functional correctness at the RTL level, we used the Synopsys Design Compiler and an open-source 15nm technology node library \cite{martins:open} to extract power, area and timing estimates from our design. 
We find that the \sys pipeline can sustain the target line rate of 100Gbps with a 256-bit interface and a cycle time of 2.56ns (i.e.,  $\sim$400MHz frequency).
The \sys pipeline adds only 3 cycles on the ingress path and 2 cycles on the egress path, for a total of  12.8ns on each packet's roundtrip time.
The design accounts for an area of 6.4$mm^2$ and 2.4 Watts of peak power, dominated by active power. 

Scaling the \sys pipeline's capabilities for higher line rates (200Gbps+) is straightforward, via two design knobs: frequency and interface width.
Given our design's low frequency of 400MHz, scaling it by $3-4\times$ is possible without microarchitectural changes.
For a given frequency, increasing the interface width from 256 to 512 bits can double \sys's sustainable line rate, without bearing significant microarchitectural restructuring of the pipeline either.
For higher line rates, SRAM buffering structures need to scale to accommodate the higher bandwidth-delay product.
Due to \sys's logic simplicity, the pipelines' area is dominated by the provisioned SRAM and hence scales almost linearly with it.
Area is typically not a major concern for modern NICs, which require large dies to drive enough pins in support of growing line rates, resulting in sparse silicon resource utilization on the die.

}
\section{Discussion}
\label{sec:discussion}

\noindent\textbf{New interfaces.}
Our microarchitectural analysis found the PCIe I/O interface as the main bottleneck {for servers  handling shallow NFs at high network bandwidth. The limits of PCIe under high DMA rates have been thoroughly studied before \cite{neugebauer:understanding}}. Even if new off-chip interfaces ameliorate PCIe limitations, bottlenecks due to intensive data movement may emerge in other system components. Our experiment with DDIO disabled demonstrated that increased queuing in memory can have detrimental effects, an effect that will only grow with increasing network line rates \cite{sutherland:nebula}. By eliminating unnecessary data movement, \sys ameliorates any on-server bottlenecks associated with data deluge, regardless of the specific component they are exhibited on.

\noindent\textbf{Limitations.} 
\sys stores sliced packet payloads in on-NIC memory resources.
As this payload placement requires packets processed by the NFs to be transmitted by the same NIC they were received from, the technique is not directly applicable to multi-NIC servers.
A possible extension could apply to dual-NIC servers that apply a known packet steering pattern, receiving packets on NIC A and transmitting them on NIC B.
In such case,  dedicated direct inter-NIC connectivity (e.g., Mellanox Socket Direct \cite{mellanox:socket-direct}) would allow NIC A to forward packet payloads to NIC B without utilizing on-server resources for such transfers.

\noindent\textbf{Additional opportunities.} 
Our experiments in \cref{sec:eval:uarch} demonstrated that \sys's benefits grow further when DDIO is disabled, because of increased latency and contention exposed by data transfers from/to memory.
We expect to see an even more pronounced effect if the network ring buffers used by the NFs are mapped on a remote socket, as inter-socket links introduce considerable latency and bandwidth limitations \cite{david:everything}.

In addition to its performance gains, we expect \sys to also deliver a secondary benefit in terms of energy reduction. \sys drastically reduces data movement over PCIe (\cref{fig:pcieReadsWrites}), which should more than offset the NIC's power draw increase due to the introduction of Slice \& Splice operations.
Data movement in general dominates energy consumption, and off-chip interfaces are particularly energy-hungry: transferring data over an off-chip interface consumes 1--2 orders of magnitude more energy than accessing a local memory structure (i.e., a cache) \cite{dally:challenges,horowitz:computing}.
{An accurate energy reduction evaluation would require end-to-end measurement including a hardware \sys implementation. To get a first-order estimate, we measured our entire NFServer's power draw as a function of packet size. 
At a 7Mpps processing rate, power draw drops from 230 to 221 Watts when the packet size is reduced from 1518B to 64B.}

\section{Related Work}
\label{sec:relwork}

\paragraph{Network data movement optimization}
As growing network line rates are gradually approaching data transfer rates conventionally exclusive to memory systems, on-server movement of network-injected data can drastically affect the memory hierarchy’s---and, by extension, the whole system’s---performance. Sutherland \etal{} demonstrated the negative performance impact such memory bandwidth interference can have on future systems, arguing for more sophisticated network data movement policies within the memory hierarchy  \cite{sutherland:nebula}. 
A body of recent work focuses on optimizing network data placement in the last-level cache, addressing performance bumps of default DDIO behavior, such as the ``leaky DMA'' problem \cite{farshin:reexamining, tootoonchian:resq, yuan:forget}.
CacheDirector implements intelligent data placement policies to place each packet header in the LLC slice closest to the core processing it \cite{farshin:make}.
Instead of optimizing data movement within the server's memory hierarchy, \sys directly decreases the volume of data moved on/off the server and within its memory hierarchy.

\paragraph{NF frameworks}
The growing popularity of NF consolidation on general-purpose servers has given rise to userspace packet processing frameworks which provide ease of high-performance NF application development. NetVM \cite{hwang:netvm} 
employs VMs to provide function isolation and shared pages for facilitating inter-VM communication. Netbricks \cite{panda:netbricks} improves upon this with containers and  language-enforced static checks to ensure packet isolation. 
Sadok \etal{}~\cite{sadok:case} distribute packets to cores at a packet rather than flow granularity to achieve even load distribution. 
Parabox \cite{zhang:parabox} and NFP \cite{sun:nfp} improve NF chain processing scalability via parallel and distributed processing. 
\sys's focus is orthogonal to these NF processing frameworks and can be combined with them to accelerate shallow NF processing of large packets.

\paragraph{Hardware offloading}
Programmable (RMT) switches has seen wide applicability in offloading computation in recent years.
PayloadPark \cite{goswami:parking}, which we have already extensively discussed, leverages RMT switches for storing payloads.
Programmable switches have also been leveraged to accelerate other network-intensive applications, such as key-value stores \cite{jin:netcache, liu:incbricks}.
Other prior work in this domain has improved application performance of key-value stores \cite{li:kv-direct}; improved NF performance by increasing throughput and reducing latency through FPGA offloads \cite{li:clicknp}; and utilized GPUs for NF acceleration \cite{han:packetshader, kim:nba, zhang:g-net}.
\sys aims to improve the performance of general-purpose servers handling NFs, which have gained traction over specialized middleboxes due to their flexibility, ease of programmability, and wide accessibility.

\paragraph{Advanced and reconfigurable NICs}
As general-purpose logic is running out of steam, hardware specialization is picking up, and is evidenced in growing capabilities of modern NICs, taking the form of advanced offloads or increased programmability. Hardened IP blocks implementing basic networking operations such as checksums or encryption are already commercialized in modern NICs, and there is a strong wave for offloading more advanced functionality. For instance, recent work demonstrates L5 protocol processing offloading  without having to migrate the L2--L4 stack onto the NIC \cite{pismenny:autonomous}. Dagger \cite{lazarev:dagger} overcomes the limitations of the PCIe interface by leveraging memory interconnects and alleviates software overheads of the RPC stack by offloading it onto an FPGA-based NIC. Ibanez \etal{} \cite{ibanez:nanopu} demonstrate improvements in nano-second scale RPCs by completely bypassing the memory and cache hierarchy with a NIC-CPU co-design. NICA \cite{eran:nica} introduces a framework that expands the SmartNIC capabilities of {inline processing of application traffic} to multiple tenants.
{A significant body of work proposes new NIC architectures \cite{kaufmann:flexnic, lin:panic}, while there have already been several successful attempts of offloading critical applications or higher-level networking functionality to advanced programmable NICs in production environments \cite{caulfield:cloud-scale, firestone:accelnet, putnam:reconfigurable}.

\sys introduces the Slice \& Splice operation as a basic building block and data movement optimization that can be leveraged by any shallow NF.
We note that the operation could be implemented in modern programmable NICs that offer sufficient resources to sustain \sys's functionality at line rate. 
However, dedicating programmable resources on the NIC to implement \sys is both unnecessary and wasteful, as the required functionality involves mostly storage and very simple logic, making it a great fit for a hardened IP block.

Finally, for systems featuring programmable NICs, the Slice \& Splice operation represents a new distinct component that an NF deployment framework \cite{katsikas:metron} may determine to offload to the NIC instead of entire NFs, as it is often infeasible to fully offload all NFs in multi-tenant environments \cite{li:dhl}.
We believe that implementing the Slice \& Splice as a hardware-accelerated operation in the NIC is a more resource-efficient option usable by multiple NFs than offloading entire NFs.
}
\section{Conclusion}
\label{sec:conclusion}

This paper demonstrated that data movement is a first-order performance determinant for NF processing of large network packets.
As shallow NFs only operate on packet headers, we identified the opportunity to directly mitigate the overhead of redundant data movement.
We introduced a packet Slice \& Splice operation on the NIC to reduce NIC-server data movement to only the small portion of each packet that is needed by the shallow NFs executing on the server.
We developed an \sys emulation platform and showed an improvement on the median and 90th percentile latency by 17--20\% and 9--29\%, respectively, for a range of shallow NFs. We further showed that for higher packet rates exceeding our emulation platform capabilities, the tail latency improvement potential grows to 55\%.
Finally, our hardware synthesis results showcased the practical feasibility of an \sys implementation as a hardware IP block 
on next-generation NICs.
\balance
\bibliographystyle{plain}

\bibliography{gen-abbrev,dblp,references}

\end{document}